\documentclass{lmcs}


\usepackage{amssymb}
\usepackage{booktabs,array}
\usepackage{pgfplots}
\pgfplotsset{compat=1.18}
\usetikzlibrary{arrows.meta}

\newcommand{\Bstar}{B_*}
\newcommand{\Rstar}{\rho_*}
\newcommand{\Farey}{\mathcal{F}}
\newcommand{\Lassos}{\mathcal{L}}
\newcommand{\Intervals}{\mathcal{I}}
\newcommand{\Obs}{\mathcal{O}}
\newcommand{\cost}{\operatorname{cost}}
\newcommand{\capreq}{\operatorname{cap}}
\newcommand{\ind}{\mathbf{1}}

\begin{document}

\title[Rate--burst safety languages]{Rate--burst safety languages: the Farey structure of the burst boundary and discrimination under erasures}

\author[L.~Dong]{Lihui Dong\lmcsorcid{0000-0002-7509-0974}}[a]
\author[Z.~Li]{Zhuo Li\lmcsorcid{0009-0005-5465-4071}}[b]

\address{Xi'an Precision Machinery Research Institute Kunming Branch, Kunming, China}
\email{dlh1997999@163.com}

\address{Institute of Scientific and Technical Information of China, Beijing, China}
\email{liz@istic.ac.cn}

\begin{abstract}
\noindent A leaky-bucket constraint with rate \(\rho\) and burst \(B\) allows at most \(\rho n+B\) costly steps in every window of \(n\) consecutive steps. The infinite 0/1 words that satisfy it form a safety language. For a finite graph with edge costs 0 and 1, we study the least burst at which the cost word of some infinite path lies in this language, as a function of the real rate. This function is the exact tolerance of a monitor that distinguishes two finite-state systems observed through an erasure channel with a leaky-bucket bound. The language is \(\omega\)-regular at every rational rate but not at irrational rates once the burst reaches 1, so ultimately periodic witnesses do not come from automata theory. We prove directly that simple lassos suffice at all real rates and that all breakpoints of the boundary lie in the Farey sequence of order N for a graph with N vertices. One lasso has a tight bound of order N to the power 2/3 on its affine pieces, whereas the choice among paths creates downward jumps and boundaries that are neither convex nor concave. One evaluation at the mediant of each Farey interval recovers the affine piece and an optimal lasso, and a greatest-fixed-point computation over decreasing integer bursts yields the whole function in polynomial time and linear working space. We also compute discrimination horizons and extend the Farey structure to turn-based safety games.
\end{abstract}

\maketitle

\section{Introduction}\label{sec:intro}

A rate--burst constraint bounds the number of events in every time window by an affine function of the window length. In network calculus, a flow regulated by a leaky bucket with rate \(\rho\) and burst \(B\) sends at most \(\rho n+B\) units in any \(n\) consecutive slots \cite{cruz1991,leboudec2001,bouillard2008}. This paper imposes the constraint on the costs along infinite paths of a finite graph. Let \(G=(V,E,S,c)\) be a directed graph with \(N\) vertices, \(M\) edges, a set \(S\subseteq V\) of initial vertices, and edge costs \(c:E\to\{0,1\}\). An infinite path \(e_1e_2\cdots\) from \(S\) is \emph{\((\rho,B)\)-feasible} if
\begin{equation}\label{eq:budget}
\sum_{t=s+1}^{u}c(e_t)\le\rho(u-s)+B\qquad\text{for all }0\le s<u.
\end{equation}
The central question of this paper is the following: for every real rate \(\rho\in[0,1]\), what is the least burst \(\Bstar(\rho)\) for which a \((\rho,B)\)-feasible infinite path exists, and how can the function \(\Bstar\) be described and computed on the whole interval \([0,1]\)?

The question comes from mode discrimination under lossy observation. A monitor observes a system that runs in one of two modes, each given by a finite automaton, and must decide which mode is active. The outputs reach the monitor through a clocked erasure channel: every scheduled observation arrives, but some arrive as an erasure symbol. Existing models of lossy observation bound the number of consecutive losses \cite{oliveira2020,oliveira2022} or the set of events whose observations may be lost \cite{carvalho2012}. A leaky-bucket bound instead allows infinitely many erasures, provided that every window of \(n\) observations contains at most \(\rho n+B\) of them. Section~\ref{sec:discrimination} shows that the monitor reaches a decision in finite time if and only if \(B<\Bstar(\rho)\) for the comparison graph of the two modes, the product of the two automata in which an edge has cost \(1\) when the two outputs differ. The function \(\Bstar\) is therefore the exact tolerance profile of the monitor: for each sustained erasure rate, it gives the burst at which the two modes can first remain confused forever.

At a fixed rate, feasibility is a one-dimensional energy problem with a weak upper bound: the energy \(E_t=\min\{B,E_{t-1}+\rho-c(e_t)\}\) starts at \(E_0=B\), energy above \(B\) is discarded, and the energy must remain nonnegative. Polynomial algorithms decide this problem at one rate \cite{bouyer2008,helouet2022}, and at a rational rate the least burst can be found by binary search over integer capacities. Such rate-by-rate methods do not show where the optimal path changes, how many affine pieces \(\Bstar\) has, or whether finitely many evaluations determine \(\Bstar\) at all real rates. Parametric shortest paths do not answer these questions either. Shortest-path distances, as functions of a parameter, are lower envelopes of affine functions, hence concave and continuous where finite \cite{karp1981,young1991}. The burst objective first maximizes over windows and then minimizes over infinite paths. The resulting boundary can jump downward, and even a continuous boundary can be neither convex nor concave (Proposition~\ref{prop:continuous}).

The constraint also has a direct reading in the theory of infinite words. The infinite 0/1 words that satisfy \eqref{eq:budget} form a set \(W(\rho,B)\) that is closed in the Cantor topology, so the constraint is a safety property, and \(\Bstar(\rho)\) is the least \(B\) for which the cost word of some infinite path of \(G\) lies in \(W(\rho,B)\). Computing \(\Bstar\) on \([0,1]\) is therefore a parameter synthesis problem: it describes all pairs \((\rho,B)\) for which a finite system meets an existential safety requirement. The rate is the delicate parameter. At a rational rate \(p/q\), a deterministic automaton that counts the scaled load accepts \(W(p/q,B)\), and the emptiness check for \(\omega\)-regular languages returns an ultimately periodic witness. At an irrational rate with \(B\ge1\), the set \(W(\rho,B)\) is not \(\omega\)-regular (Theorem~\ref{thm:regular}). This route is then closed, and we prove the existence of lasso witnesses directly. The whole boundary is nevertheless determined by finitely many rational rates, one of denominator at most \(2N\) in each Farey interval.

\begin{figure}[t]
\centering
\begin{tikzpicture}[>=Stealth,vertex/.style={circle,draw,minimum size=7mm,inner sep=1pt},every node/.style={font=\small}]
\node[vertex] (h) at (0,0) {\(h\)};
\node[vertex] (o1) at (1.5,1.2) {\(o_1\)};
\node[vertex] (o2) at (3.0,1.2) {\(o_2\)};
\node[vertex] (y2) at (3.0,-1.2) {\(y_2\)};
\node[vertex] (y1) at (1.5,-1.2) {\(y_1\)};
\draw[->] (-0.9,0.8)--(h);
\draw[->] (h) to[bend left=18] node[above left]{\(1\)} (o1);
\draw[->] (o1) to[bend left=18] node[below right]{\(0\)} (h);
\draw[->] (o1)--node[above]{\(1\)}(o2);
\draw[->] (o2)--node[right]{\(0\)}(y2);
\draw[->] (y2)--node[below]{\(0\)}(y1);
\draw[->] (y1)--node[below left]{\(0\)}(h);
\end{tikzpicture}
\hspace{1cm}
\begin{tikzpicture}
\begin{axis}[width=7cm,height=5cm,xmin=0,xmax=1.02,ymin=0,ymax=1.45,xlabel={Rate \(\rho\)},ylabel={\(\Bstar(\rho)\)},xtick={0,0.4,0.5,1},xticklabels={\(0\),\llap{\(2/5\)},\rlap{\(1/2\)},\(1\)},ytick={0,0.5,1,1.2},yticklabels={\(0\),\(1/2\),\(1\),\(6/5\)},axis lines=left,tick label style={font=\footnotesize},label style={font=\small},clip=false]
\addplot[draw=none,fill=teal!12] coordinates {(0.4,1.2)(0.5,1)(0.5,1.4)(0.4,1.4)} \closedcycle;
\addplot[draw=none,fill=teal!12] coordinates {(0.5,0.5)(1,0)(1,1.4)(0.5,1.4)} \closedcycle;
\addplot[very thick,teal!70!black,domain=0.4:0.5]{2-2*x};
\addplot[very thick,teal!70!black,domain=0.5:1]{1-x};
\addplot[only marks,mark=*,mark size=2pt,teal!70!black] coordinates {(0.4,1.2)(0.5,0.5)(1,0)};
\addplot[only marks,mark=*,mark options={fill=white,draw=teal!70!black},mark size=2.2pt] coordinates {(0.5,1)};
\draw[dashed,gray] (axis cs:0.4,0)--(axis cs:0.4,1.4);
\end{axis}
\end{tikzpicture}
\caption{Left: a graph with initial vertex \(h\) and two simple cycles, \(h o_1 h\) with cost word \(10\) and \(h o_1 o_2 y_2 y_1 h\) with cost word \(11000\). Right: its burst boundary, which is infinite on \([0,2/5)\), equals \(2-2\rho\) on \([2/5,1/2)\), and equals \(1-\rho\) on \([1/2,1]\). Shading marks the pairs \((\rho,B)\) that admit a feasible infinite path. The open point is excluded.}
\label{fig:example}
\end{figure}
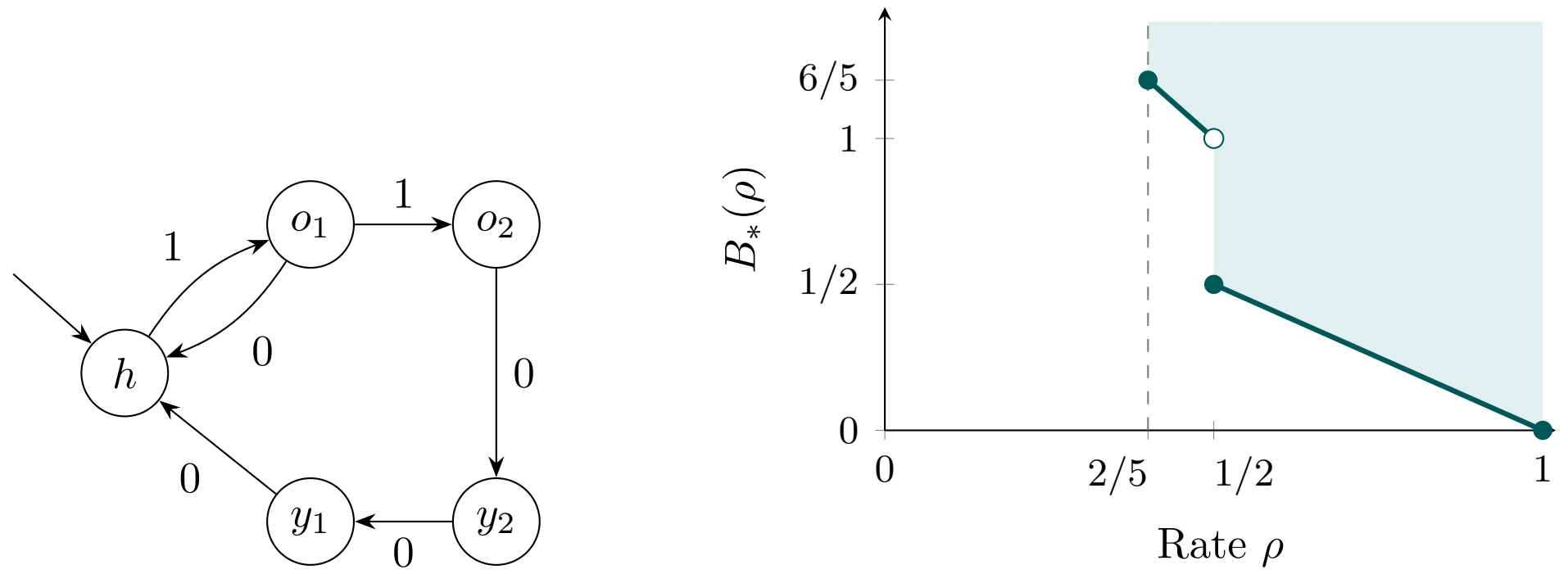

Figure~\ref{fig:example} shows a small example. The graph has two cycles through its initial vertex \(h\). Reading the edge costs along them gives the cost words \(10\) and \(11000\), with mean costs \(1/2\) and \(2/5\). For \(\rho<2/5\), each cycle costs more than \(\rho\) times its length, so long windows violate \eqref{eq:budget} for every finite \(B\). For \(2/5\le\rho<1/2\), a feasible path must traverse the cycle \(11000\) infinitely often, and its two consecutive cost-\(1\) edges force \(B\ge2-2\rho\); the path that repeats this cycle attains this bound. At \(\rho=1/2\), the cycle \(10\) becomes usable, which means that its mean cost no longer exceeds \(\rho\), and it needs only \(B=1-\rho\). The boundary thus drops from \(1\) to \(1/2\) at \(\rho=1/2\). In the discrimination reading, the edges of cost \(1\) are warnings emitted by one mode while the other mode emits ordinary reports, and the monitor tolerates a burst just below \(2-2\rho\) when \(2/5\le\rho<1/2\) but only a burst below \(1-\rho\) when \(\rho\ge1/2\).

\subsection*{Contributions}

Our results separate the geometry of one path from the choices available in a graph, and Table~\ref{tab:results} lists where each result is proved.
\begin{enumerate}
\item The feasible words form a safety language that is \(\omega\)-regular at every rational rate and is not \(\omega\)-regular at irrational rates once \(B\ge1\). At every real rate, whenever a feasible infinite path exists, a feasible simple lasso exists. The burst requirement of a simple lasso, where finite, is a maximum of lines \(k-\ell\rho\) with integers \(0\le k\le\ell\le N-1\). A single simple lasso has \(\Theta(N^{2/3})\) affine pieces in the worst case, and a deterministic path ending in a zero-cost self-loop attains this order.
\item The burst boundary of a graph has \(O(N^2)\) affine pieces, and all its breakpoints lie in the Farey sequence of order \(N\). We characterize when a jump occurs, prove that \(\Bstar(\rho)/(1-\rho)\) is nonincreasing, and show that the last piece is \(R(1-\rho)\), where \(R\) is the least possible length of the longest run of cost-\(1\) edges on an infinite path. A family of graphs has \((N+1)/3\) finite pieces and \((N-2)/3\) downward jumps. The critical-rate function, which gives the least rate at a fixed burst, is continuous.
\item The value of \(\Bstar\) at the mediant of a Farey interval determines the affine piece on that interval and a simple lasso that is optimal on it. A sweep over decreasing integer capacities with one load threshold per vertex computes the whole boundary in \(O((N+M)N^4)\) arithmetic operations and \(O(N+M)\) working space.
\item Below the minimum cycle mean \(\mu\), the discrimination horizon is finite for every burst, can be computed exactly at rational parameters, and grows as \(B/(\mu-\rho)\).
\item In fully observed turn-based safety games with the same constraint, the least burst has the same Farey structure, and one positional strategy is optimal throughout each Farey interval. A Büchi condition destroys right continuity, even in a one-vertex arena.
\end{enumerate}

\begin{table}[t]
\centering
\caption{Main results. \(N\) and \(M\) are the numbers of vertices and edges of the graph or arena.}
\label{tab:results}
\small
\begin{tabular}{@{}>{\raggedright\arraybackslash}p{0.2\linewidth}>{\raggedright\arraybackslash}p{0.5\linewidth}>{\raggedright\arraybackslash}p{0.24\linewidth}@{}}
\toprule
Question & Result & Location\\
\midrule
Feasible words & Safety language; \(\omega\)-regular at rational rates, not at irrational rates when \(B\ge1\) & Theorem~\ref{thm:regular}\\
Witness paths & Simple lassos suffice at all real rates & Theorem~\ref{thm:lasso}\\
One lasso & \(\Theta(N^{2/3})\) pieces in the worst case & Theorem~\ref{thm:singlelasso}\\
Whole boundary & \(O(N^2)\) pieces, breakpoints in \(\Farey_N\), characterization of jumps, terminal piece \(R(1-\rho)\) & Theorem~\ref{thm:frontier}, Propositions~\ref{prop:jumps}, \ref{prop:terminal}\\
Shape & Continuous examples that are neither convex nor concave; \((N-2)/3\) jumps & Propositions~\ref{prop:continuous}, \ref{prop:manypieces}\\
Fixed burst & Continuous critical rate with slopes in \(\{0\}\cup\{-1/\ell\}\) & Theorem~\ref{thm:inverse}\\
Computation & \(O((N+M)N^4)\) operations, \(O(N+M)\) working space & Theorem~\ref{thm:algorithm}\\
Discrimination time & Exact value at rational parameters; growth \(1/(\mu-\rho)\) & Proposition~\ref{prop:horizon}, Theorem~\ref{thm:delay}\\
Safety games & Same Farey structure; one positional strategy per interval & Theorems~\ref{thm:gamefrontier}, \ref{thm:gamesynthesis}\\
Büchi acceptance & Right continuity fails & Proposition~\ref{prop:buchi}\\
\bottomrule
\end{tabular}
\end{table}

The algorithm rests on two observations. At a Farey mediant \(p/q\), the integer \(q\Bstar(p/q)=qk-p\ell\) determines the coefficients \((k,\ell)\), because \(\ell<q\). In the product of \(G\) with the load automaton, the states from which an infinite path starts form the greatest fixed point of the predecessor operator, as for every safety objective, and at each vertex their loads form an interval starting at \(0\). One integer per vertex therefore represents this fixed point, and maintaining these integers while the capacity decreases finds the least feasible capacity in one pass.

Section~\ref{sec:related} reviews related work. Section~\ref{sec:boundary} defines the safety language of feasible words and the burst boundary and determines when the language is \(\omega\)-regular, and Section~\ref{sec:discrimination} derives it from the discrimination problem. Section~\ref{sec:lassos} proves the lasso reduction and the sharp piece bound for one lasso. Section~\ref{sec:frontier} studies the boundary of a graph and the critical-rate function, and Section~\ref{sec:algorithm} gives the algorithm. Section~\ref{sec:delay} computes discrimination horizons, Section~\ref{sec:games} treats safety games, and Section~\ref{sec:conclusion} concludes.

\section{Related work}\label{sec:related}

\paragraph{Diagnosis under unreliable observations.} Diagnosability of discrete-event systems was formalized by Sampath et al.~\cite{sampath1995}, and Cassandras and Lafortune give a textbook account \cite{cassandras2021}. Several works relax the assumption that every observable event reaches the diagnoser. Athanasopoulou et al.~\cite{athanasopoulou2010} compute the most likely failure mode when sensors insert, delete, or transpose symbols with known probabilities. Carvalho et al.~\cite{carvalho2012} model intermittent loss of observations by language dilation, and Oliveira et al.~\cite{oliveira2020,oliveira2022} bound the number of consecutive losses. Lin~\cite{lin2014} and Nunes et al.~\cite{nunes2018} treat communication delays and losses in networked supervisory control and codiagnosis. Dong et al.~\cite{dong2024} specify unreliable sensors in linear temporal logic and use a compactness argument similar to Proposition~\ref{prop:compact}. Lin et al.~\cite{lin2024} and Li et al.~\cite{li2025} analyze diagnosis under sensor attacks with deletions, insertions, and substitutions. These works decide diagnosability for one given loss model. In our model the erasure symbol preserves the observation clock, the number of erasures may be unbounded, and we compute the set of all rate--burst pairs under which the modes can be distinguished in finite time.

\paragraph{Network calculus.} The window constraint \eqref{eq:budget} is the affine arrival curve of network calculus, which describes traffic regulated by a leaky bucket \cite{cruz1991,leboudec2001,bouillard2008}. Zippo et al.~\cite{zippo2023} state it in the discrete-time form used here and give algorithms for ultimately pseudo-periodic functions. In network calculus the rate and the burst describe a given flow. Here they are optimized over the infinite paths of a graph and, in Section~\ref{sec:games}, over the moves of an adversary.

\paragraph{Energy problems.} Energy problems were introduced by Chakrabarti et al.~\cite{chakrabarti2003}, and Brim et al.~\cite{brim2011} compute the least initial credit by a fixpoint iteration. Bouyer et al.~\cite{bouyer2008} decide the existence of an infinite run under a weak upper bound in polynomial time for integer weights and prove that memoryless strategies suffice in the one-dimensional case. Hélouët et al.~\cite{helouet2022} study reachability games with relaxed energy constraints and show that, if some weak upper bound admits an infinite run, then one of size at most \(N\) times the largest absolute weight does. Juhl et al.~\cite{juhl2013} treat the upper bound, or the weights of some transitions, as parameters and construct the set of admissible values from its minimal elements. Dziadek et al.~\cite{dziadek2025} solve energy problems with Büchi and parity acceptance conditions. Farey sequences and reweighting also appear in mean-payoff algorithms \cite{comin2017}. All these methods work at one rate at a time. Our parameter is the rate, which is subtracted from every edge weight, and we describe the least burst as a function of a real rate.

\paragraph{Parametric optimization.} When every edge length has a parameter coefficient in \(\{0,-1\}\), shortest-path distances, as functions of the parameter, have polynomially many pieces \cite{karp1981,young1991}; superpolynomial lower bounds concern unrestricted affine coefficients \cite{carstensen1983,mulmuley2001}. Shortest-path values are concave where finite. For a single lasso, our piece bound uses the lattice-point count of Jarník \cite{jarnik1926} for convex curves.

\paragraph{Quantitative languages and parameter synthesis.} Quantitative languages assign a value to every infinite word, and Chatterjee et al.~\cite{chatterjee2010} compare the supremum, limit-average, and discounted values computed by weighted automata. For a fixed rate, the least burst of a word, \(\sup_{s<u}(\sum_{t=s+1}^{u}z_t-\rho(u-s))\), is such a value, and \(\Bstar(\rho)\) is the least value of a cost word of \(G\). This value is a supremum over all windows rather than over prefixes, and it depends on a real parameter. Parameter synthesis asks for all parameter values under which a system satisfies a property; for parametric timed automata, already the emptiness problem is undecidable \cite{alur1993}. In our setting the parameters enter linearly into a one-counter safety condition, and the set of feasible pairs \((\rho,B)\) is the epigraph of a piecewise affine function with Farey breakpoints, which need not be convex. Mechanical words \cite{lothaire2002} have the most even distribution of ones for a given frequency. At an irrational rate they belong to \(W(\rho,1)\), and this is the source of nonregularity in Theorem~\ref{thm:regular}.

\paragraph{Window objectives.} Window objectives bound the time within which accumulated payoff must recover. Chatterjee et al.~\cite{chatterjee2015} introduce window mean-payoff objectives for games, and Ji et al.~\cite{ji2022} synthesize supervisors for local mean-payoff conditions. Benedikt et al.~\cite{benedikt2014} compute the asymptotic per-character cost of repairing regular languages and relate streaming repair to mean-payoff games. Our constraint has no window that must close: inequality \eqref{eq:budget} must hold for every window, of every length and at every position. Proposition~\ref{prop:buchi} shows how a Büchi requirement changes the resulting boundary.

\section{Rate--burst safety languages and the burst boundary}\label{sec:boundary}

We first state the constraint for words and then for paths. For a finite or infinite word \(z=z_1z_2\cdots\) over \(\{0,1\}\), a \emph{window} is a factor \(z_{s+1}\cdots z_u\) with \(s<u\), and its \emph{excess} at rate \(\rho\) is \(\sum_{t=s+1}^{u}z_t-\rho(u-s)\). The word is \((\rho,B)\)-feasible if every window has excess at most \(B\), that is, if it satisfies \eqref{eq:budget} with \(z_t\) in place of \(c(e_t)\). Let
\begin{equation}\label{eq:language}
W(\rho,B)=\{z\in\{0,1\}^\omega:z\text{ is }(\rho,B)\text{-feasible}\}.
\end{equation}
Every window of an infinite word lies in a finite prefix, so an infinite word belongs to \(W(\rho,B)\) exactly when all its finite prefixes are feasible. In the Cantor topology on \(\{0,1\}^\omega\), this means that \(W(\rho,B)\) is closed, that is, a safety property \cite{perrin2004}.

Fix \(G=(V,E,S,c)\) as in Section~\ref{sec:intro}, with \(N=|V|\ge1\), \(M=|E|\), and \(S\ne\varnothing\). We assume that every vertex is reachable from \(S\), and we allow parallel edges. Paths start in \(S\) unless another start vertex is named. For a finite path \(P\), let \(|P|\) be its number of edges and \(\cost(P)\) its total cost. The cost word of a path \(e_1e_2\cdots\) is \(c(e_1)c(e_2)\cdots\), and a path is feasible if its cost word is feasible. Let \(L(G)\subseteq\{0,1\}^\omega\) be the set of cost words of infinite paths.

Along a word, or along the cost word of a path, the reflected load is
\begin{equation}\label{eq:load}
x_0=0,\qquad x_t=\max\{0,x_{t-1}+c(e_t)-\rho\}.
\end{equation}
Induction on \(t\) shows that
\begin{equation}\label{eq:loadexpansion}
x_t=\max_{0\le s\le t}\Bigl(\sum_{j=s+1}^{t}c(e_j)-\rho(t-s)\Bigr),
\end{equation}
where the empty sum at \(s=t\) is zero. Thus \(x_t\) is the largest excess of a window ending at time \(t\), or \(0\) if all these excesses are negative, and a path is feasible exactly when \(x_t\le B\) for all \(t\). In network calculus, \(x_t\) is the backlog of a virtual queue served at rate \(\rho\). The energy \(E_t=B-x_t\) evolves by
\begin{equation}\label{eq:energy}
E_0=B,\qquad E_t=T_{e_t}(E_{t-1}),\qquad T_e(\eta)=\min\{B,\eta+\rho-c(e)\},
\end{equation}
and must remain nonnegative. This is the energy model with a weak upper bound studied in \cite{bouyer2008,helouet2022,dziadek2025}: energy above the capacity \(B\) is discarded.

\begin{defi}\label{def:frontier}
The burst boundary of \(G\) is the function \(\Bstar:[0,1]\to[0,\infty]\) defined by
\begin{equation}\label{eq:defBstar}
\Bstar(\rho)=\inf\{B\ge0:L(G)\cap W(\rho,B)\ne\varnothing\},
\end{equation}
with \(\inf\varnothing=\infty\). Equivalently, \(\Bstar(\rho)\) is the infimum of the bursts \(B\) for which a \((\rho,B)\)-feasible infinite path exists.
\end{defi}

The restriction \(\rho\le1\) loses nothing: at \(\rho=1\) every infinite path is feasible with \(B=0\), so \(\Bstar(\rho)\) is \(0\) for \(\rho\ge1\) whenever an infinite path exists.

\begin{lem}\label{lem:compact}
For every \(\rho\in[0,1]\) and \(B\ge0\), \(G\) has a feasible infinite path if and only if it has feasible finite paths of every length.
\end{lem}

\begin{proof}
Feasible finite paths form a prefix-closed, finitely branching tree. If its depth is unbounded, König's lemma gives an infinite branch, and every window of that branch lies in a feasible prefix. The converse follows by taking prefixes.
\end{proof}

The argument uses neither rationality of \(\rho\) nor integrality of \(B\).

\subsection{When the feasible words form an \texorpdfstring{\(\omega\)}{omega}-regular language}\label{sec:regular}

A deterministic safety automaton over \(\{0,1\}\) is a triple \(D=(Q,q_0,\delta)\) consisting of a finite set \(Q\) of states, an initial state \(q_0\), and a partial transition function \(\delta:Q\times\{0,1\}\to Q\). It accepts an infinite word if its run from \(q_0\) is defined at every step. Every language accepted by such an automaton is \(\omega\)-regular and closed \cite{perrin2004}. The next theorem shows that \(W(\rho,B)\) has this form at every rational rate, and that at an irrational rate with \(B\ge1\) it is not even \(\omega\)-regular.

\begin{thm}\label{thm:regular}
Let \(B\ge0\).
\begin{enumerate}
\item If \(\rho=p/q\) with \(0\le p\le q\) and \(\gcd(p,q)=1\), let \(K=\lfloor qB\rfloor\). Then \(W(\rho,B)\) is accepted by the deterministic safety automaton \(D_{p,q,K}\) with states \(\{0,\ldots,K\}\), initial state \(0\), and transitions
\begin{equation}\label{eq:loadautomaton}
\delta(j,a)=\max\{0,j+qa-p\}\qquad\text{whenever this value is at most }K.
\end{equation}
\item If \(\rho\in(0,1)\) is irrational and \(B\ge1\), then \(W(\rho,B)\) is not \(\omega\)-regular.
\end{enumerate}
\end{thm}

\begin{proof}
For (1), compare \eqref{eq:loadautomaton} with \eqref{eq:load}: after reading \(t\) letters, the automaton is in state \(qx_t\) as long as all earlier states are at most \(K\). The scaled load \(qx_t\) is an integer, so \(x_t\le B\) holds exactly when \(qx_t\le K\). The run is therefore defined forever exactly when \(x_t\le B\) for all \(t\), that is, when the word is feasible.

For (2), consider the mechanical word \(m_t=\lfloor t\rho\rfloor-\lfloor(t-1)\rho\rfloor\), \(t\ge1\). Its letters are \(0\) or \(1\) because \(0<\rho<1\), and every window satisfies
\begin{equation}\label{eq:mechanical}
\sum_{t=s+1}^{u}m_t=\lfloor u\rho\rfloor-\lfloor s\rho\rfloor<\rho(u-s)+1,
\end{equation}
so \(m\in W(\rho,B)\). Suppose that \(W(\rho,B)\) is \(\omega\)-regular. Its set \(P\) of finite prefixes is then a regular language \cite{perrin2004}. Take a deterministic finite automaton for \(P\), keep only the states reached by words of \(P\), and delete the transitions that leave this set. In the resulting deterministic safety automaton \(D\), every state has a successor, because every word of \(P\) extends by one letter to a word of \(P\). The run of \(D\) on an infinite word is defined forever exactly when all prefixes lie in \(P\), so \(D\) accepts the closed set \(W(\rho,B)\).

Regard \(D\) as a graph whose edges carry the letters as costs, let \(n\) be its number of states, and let \(m^*\) be the largest mean cost of a simple cycle of \(D\). The run of \(D\) on the first \(T\) letters of \(m\) decomposes into a simple path with at most \(n-1\) edges and simple cycles, so
\begin{equation}\label{eq:mechanicalcount}
\rho T-1<\lfloor\rho T\rfloor=\sum_{t=1}^{T}m_t\le n-1+m^*T\qquad(T\ge1).
\end{equation}
Letting \(T\to\infty\) gives \(m^*\ge\rho\). The mean cost of a cycle is rational and \(\rho\) is not, so \(m^*>\rho\). Let \(\beta\) be a simple cycle of \(D\) with mean cost \(m^*\), and let \(\alpha\) be a path from \(q_0\) to a state of \(\beta\). The word read along \(\alpha\beta^\omega\) is accepted by \(D\). A window consisting of \(k\) copies of the label of \(\beta\) has excess \(k|\beta|(m^*-\rho)\), which exceeds \(B\) for large \(k\). This contradicts \(D\) accepting only feasible words.
\end{proof}

Theorem~\ref{thm:regular} separates two facts that coincide for \(\omega\)-regular specifications. If \(M\subseteq\{0,1\}^\omega\) is \(\omega\)-regular, then \(L(G)\cap M\) is nonempty exactly when it contains an ultimately periodic word, and an emptiness check produces one. At a rational rate this applies to \(M=W(p/q,B)\), and the product of \(G\) with \(D_{p,q,K}\) is the load graph used in Section~\ref{sec:algorithm}. At an irrational rate the argument is unavailable, and Theorem~\ref{thm:lasso} shows that simple lassos remain sufficient. The bound \(B\ge1\) in part (2) cannot simply be dropped: if \(B<1-\rho\), no window may contain a \(1\), and \(W(\rho,B)=\{0^\omega\}\). The automaton \(D_{p,q,K}\) has \(\lfloor qB\rfloor+1\) states, which grow with the denominator \(q\); Section~\ref{sec:algorithm} evaluates \(\Bstar\) only at rates with denominator at most \(2N\).

\section{Mode discrimination under a clocked erasure channel}\label{sec:discrimination}

This section derives the burst boundary from the discrimination problem described in Section~\ref{sec:intro}. Let \(\Sigma\) be a finite output alphabet. Mode \(i\in\{0,1\}\) is a finite labelled transition system
\begin{equation}\label{eq:modes}
A_i=(Q_i,I_i,\Delta_i),\qquad \Delta_i\subseteq Q_i\times\Sigma\times Q_i,\qquad I_i\ne\varnothing.
\end{equation}
A run of length \(n\) from \(I_i\) emits a word \(a_1\cdots a_n\in\Sigma^n\). Nondeterministic transitions and several initial states are allowed. Each transition corresponds to one observation time, and the mode is fixed when observation starts.

An erasure mask of length \(n\) is a word \(z_1\cdots z_n\in\{0,1\}^n\), and the received symbol at time \(t\) is
\begin{equation}\label{eq:received}
y_t=\begin{cases}a_t,&z_t=0,\\\bot,&z_t=1,\end{cases}\qquad\bot\notin\Sigma.
\end{equation}
The receiver therefore sees which positions were erased. A channel that deletes symbols would instead hide the positions of the losses. The mask is \((\rho,B)\)-admissible if it is a \((\rho,B)\)-feasible word in the sense of Section~\ref{sec:boundary}, so the infinite admissible masks form the language \(W(\rho,B)\). For example, a block of \(k\) consecutive erasures requires \(k(1-\rho)\le B\). Every admissible finite mask extends to an admissible infinite mask by appending zeros.

Let \(\Obs_i^n(\rho,B)\) be the set of received words obtained from runs of length \(n\) of \(A_i\) under admissible masks. A received word is ambiguous if it lies in both sets.

\begin{defi}\label{def:horizon}
The discrimination horizon is
\begin{equation}\label{eq:horizon}
H(\rho,B)=\min\{n\ge1:\Obs_0^n(\rho,B)\cap\Obs_1^n(\rho,B)=\varnothing\},
\end{equation}
with \(H(\rho,B)=\infty\) if the set is empty. The modes are uniformly distinguishable under \((\rho,B)\) if \(H(\rho,B)<\infty\).
\end{defi}

Both sets of received words are prefix-closed across lengths, so once ambiguity disappears it cannot reappear at a later length. A finite \(H\) is thus a uniform guarantee: after \(H\) observations the monitor knows the mode, whatever the run and the admissible mask.

The comparison graph of \(A_0\) and \(A_1\) runs the two automata side by side. Its vertex set is the part of \(Q_0\times Q_1\) reachable from \(S=I_0\times I_1\), and each pair of transitions \((q_0,a,q_0')\in\Delta_0\) and \((q_1,b,q_1')\in\Delta_1\) gives an edge \((q_0,q_1)\to(q_0',q_1')\) with cost
\begin{equation}\label{eq:binarycost}
c(e)=\ind_{\{a\ne b\}}\in\{0,1\}.
\end{equation}
Edges with the same endpoints and the same cost are merged, which preserves all cost sequences of paths, so the comparison graph has \(M\le2N^2\) edges.

\begin{lem}\label{lem:mismatch}
The modes have an ambiguous received word of length \(n\) if and only if the comparison graph has a feasible path of length \(n\).
\end{lem}

\begin{proof}
Suppose two runs give the same received word under admissible masks \(z\) and \(z'\). Erased positions are visible because \(\bot\notin\Sigma\), so \(z=z'\), and this mask erases every position where the two outputs differ. The cost sequence of the corresponding path in the comparison graph is therefore bounded entrywise by an admissible mask. Lowering entries preserves every upper bound in \eqref{eq:budget}, so the path is feasible. Conversely, erasing exactly the positions where the outputs along a feasible path differ is an admissible mask under which the two runs give the same received word.
\end{proof}

\begin{prop}\label{prop:compact}
Let \(L(\rho,B)\in\{0,1,\ldots\}\cup\{\infty\}\) be the supremum of the lengths of feasible paths in the comparison graph. Then
\begin{equation}\label{eq:horizonL}
H(\rho,B)=L(\rho,B)+1,
\end{equation}
with \(\infty+1=\infty\). In particular, \(H(\rho,B)<\infty\) if and only if \(B<\Bstar(\rho)\).
\end{prop}

\begin{proof}
Equation \eqref{eq:horizonL} follows from Lemma~\ref{lem:mismatch} and prefix closure. By Lemma~\ref{lem:compact}, \(L=\infty\) exactly when a feasible infinite path exists, and by Corollary~\ref{cor:minlasso} below, this holds exactly when \(B\ge\Bstar(\rho)\).
\end{proof}

Every graph with 0/1 edge costs arises as a comparison graph. Let \(A_0\) have one state with a self-loop emitting \(0\), and let \(A_1\) be the graph itself with edge \(e\) emitting \(c(e)\). The comparison graph is then the original graph. All examples and lower bounds below therefore hold for discrimination as well.

The window constraint must hold at every position, not only for prefixes. The following example shows why. Consider two output sequences that agree for \(t\) steps and differ at every later step, such as \(a^\omega\) and \(a^tb^\omega\), and fix \(0<\rho<1\). Suppose the channel were constrained only on prefixes,
\begin{equation}\label{eq:prefixonly}
\sum_{j=1}^{n}z_j\le\rho n+B\qquad(n\ge1).
\end{equation}
For any block length \(m\), choose \(t\) with \(\rho t+B\ge(1-\rho)m\). Keeping the first \(t\) outputs and erasing the next \(m\) satisfies every prefix constraint, so the monitor would wait \(m\) steps after the first difference, and \(m\) is unbounded. Under \eqref{eq:budget}, the erased block itself requires \(m(1-\rho)\le B\), whatever the length of the preceding agreement. The window form thus bounds the delay after the first difference.

\section{Reduction to simple lassos}\label{sec:lassos}

A \emph{simple lasso} is an infinite path \(\lambda=\alpha\beta^\omega\) in which \(\alpha\) is a simple path from \(S\), \(\beta\) is a nonempty simple cycle, and \(\alpha\) and \(\beta\) share only the vertex where \(\beta\) starts. Then \(|\alpha|+|\beta|\le N\). Let \(\Lassos\) be the finite set of simple lassos. Its size can be exponential in \(N\).

\begin{thm}\label{thm:lasso}
For every \(\rho\in[0,1]\) and \(B\ge0\), \(G\) has a \((\rho,B)\)-feasible infinite path if and only if it has a \((\rho,B)\)-feasible simple lasso.
\end{thm}

\begin{proof}
A simple lasso is an infinite path, so only the forward direction needs proof. For a vertex \(v\), let \(W_v\subseteq[0,B]\) be the set of initial energies from which some infinite path starting at \(v\) keeps the energy \eqref{eq:energy} nonnegative. Each \(T_e\) is nondecreasing, so \(W_v\) is upward closed. For each \(h\), the set of initial energies from which some path of length \(h\) starting at \(v\) keeps the energy nonnegative is closed, because it is a finite union of closed sets and each \(T_e\) is continuous. By König's lemma, \(W_v\) is the intersection of these closed sets over all \(h\). Every nonempty \(W_v\) is therefore an interval \([\tau_v,B]\).

At each vertex \(v\) with \(W_v\ne\varnothing\), fix the first edge \(e=(v,u)\) of an infinite path that starts at \(v\) with energy \(\tau_v\), so that \(T_e(\tau_v)\ge\tau_u\). Since \(T_e\) is nondecreasing, following the fixed edges from any vertex \(v\) with energy at least \(\tau_v\) keeps the energy at least the threshold of the current vertex. Let \(s\) be the first vertex of a given feasible path. Then \(B\in W_s\), so starting at \(s\) with energy \(B\) and following the fixed edges is possible. The resulting path is feasible, and its next edge depends only on the current vertex. It is therefore a simple prefix followed by a repeated simple cycle, that is, a simple lasso.
\end{proof}

This is the memoryless-strategy argument of Bouyer et al.~\cite{bouyer2008} for energy constraints with a weak upper bound. We include it because they assume integer weights, whereas the rate and the burst are real here. By Theorem~\ref{thm:regular}, ultimately periodic witnesses of \(\omega\)-regular languages do not cover irrational rates either. Section~\ref{sec:games} extends the argument to games.

We now compute the least burst of a single simple lasso. For \(\lambda=\alpha\beta^\omega\), let \(\mu_\lambda=\cost(\beta)/|\beta|\) be its cycle mean, and let \(\Intervals_\lambda\) be the set of pairs \((k,\ell)\) such that some nonempty window of \(\alpha\beta\beta\) with at most \(N-1\) edges has cost \(k\) and length \(\ell\). Thus
\begin{equation}\label{eq:intervalpairs}
0\le k\le\ell\le N-1\qquad\text{for all }(k,\ell)\in\Intervals_\lambda.
\end{equation}
A cycle, or a lasso, is \emph{usable} at \(\rho\) if its mean cost is at most \(\rho\).

\begin{lem}\label{lem:lassocap}
Let the burst requirement \(\capreq_\lambda(\rho)\) be the least \(B\) for which \(\lambda\) is \((\rho,B)\)-feasible. Then \(\capreq_\lambda(\rho)=\infty\) for \(\rho<\mu_\lambda\), and for \(\rho\ge\mu_\lambda\),
\begin{equation}\label{eq:lassocap}
\capreq_\lambda(\rho)=\max\bigl(\{0\}\cup\{k-\ell\rho:(k,\ell)\in\Intervals_\lambda\}\bigr).
\end{equation}
\end{lem}

\begin{proof}
If \(\rho<\mu_\lambda\), each copy of \(\beta\) has positive excess, so windows containing many copies exceed every finite bound. Let \(\rho\ge\mu_\lambda\), and take a window of \(\lambda\) whose part in \(\beta^\omega\) has at least \(|\beta|\) edges. Its last \(|\beta|\) edges form a rotation of \(\beta\), whose excess \(|\beta|(\mu_\lambda-\rho)\) is nonpositive, so deleting them leaves a shorter window with at least the same excess. Repeating this step turns a window that starts in \(\beta^\omega\) into a possibly empty window of \(\beta\beta\) with fewer than \(|\beta|\) edges, and a window that starts in \(\alpha\) into a window of \(\alpha\beta\) with at most \(|\alpha|+|\beta|-1\le N-1\) edges. Both are windows of \(\alpha\beta\beta\), so every excess is at most the right-hand side of \eqref{eq:lassocap}. Every window of \(\alpha\beta\beta\) is a window of \(\lambda\), so the bound is attained.
\end{proof}

\begin{cor}\label{cor:minlasso}
For every \(\rho\in[0,1]\),
\begin{equation}\label{eq:frontiermin}
\Bstar(\rho)=\min_{\lambda\in\Lassos}\capreq_\lambda(\rho),
\end{equation}
with \(\min\varnothing=\infty\). A feasible infinite path exists exactly when \(B\ge\Bstar(\rho)\).
\end{cor}

\begin{proof}
By Lemma~\ref{lem:lassocap}, each \(\lambda\) is \((\rho,\capreq_\lambda(\rho))\)-feasible when \(\capreq_\lambda(\rho)<\infty\). By Theorem~\ref{thm:lasso}, the infimum in \eqref{eq:defBstar} is the minimum of the finitely many values \(\capreq_\lambda(\rho)\), and every finite value is attained.
\end{proof}

\subsection{The sharp complexity of one lasso}

The finite requirement of one lasso is a convex function of \(\rho\), and the coefficient points \((\ell,k)\) of its lines lie in a triangle of side length \(N-1\). Only the points that support the maximum contribute pieces, and these points form a convex lattice chain. Counting such chains as Jarník did \cite{jarnik1926} gives \(O(N^{2/3})\) pieces, instead of the \(O(N^2)\) that the Farey grid allows. For the lower bound, we turn a concave lattice polygon into a binary word whose windows can be bounded explicitly.

\begin{thm}\label{thm:singlelasso}
The finite part of the burst requirement of a single simple lasso in an \(N\)-vertex graph has \(O(N^{2/3})\) maximal affine pieces. This order is attained by deterministic graphs: for every sufficiently large \(N\), some \(N\)-vertex directed path ending in a cost-\(0\) self-loop has a burst boundary with \(\Omega(N^{2/3})\) affine pieces.
\end{thm}

\begin{proof}
For the upper bound, Lemma~\ref{lem:lassocap} expresses the finite requirement as the maximum of lines \(k-\ell\rho\) with integer coefficients \(0\le k\le\ell\le N-1\). Let \(r\) be the number of maximal affine pieces on its finite domain. Order their coefficient points \((\ell,k)\) by increasing \(\ell\). The differences between consecutive points have positive first coordinate, and their slopes are distinct numbers in \([0,1]\), because these slopes are the breakpoints of the requirement. Write each difference as a positive integer multiple of a primitive vector \((a,b)\), where \(1\le a\) and \(0\le b\le a\). The primitive vectors are distinct, and the sum of their first coordinates is at most \(N-1\). For every integer \(H\ge1\), at most \(H(H+3)/2\) of these vectors have \(a\le H\), and at most \((N-1)/(H+1)\) have \(a>H\). Consequently,
\begin{equation}\label{eq:singlelassoupper}
r\le1+\frac{H(H+3)}2+\frac{N-1}{H+1}.
\end{equation}
Taking \(H=\max\{1,\lfloor N^{1/3}\rfloor\}\) proves the upper bound.

For the lower bound, fix an integer \(m\ge1\) and take all coprime pairs \((u,v)\in\{1,\ldots,m\}^2\). Their number \(r_m\) satisfies
\begin{equation}\label{eq:coprimecount}
r_m\ge m^2-\sum_{d=2}^{m}\lfloor m/d\rfloor^2\ge\frac{m^2}{4},
\end{equation}
because \(\sum_{d=2}^{\infty}d^{-2}\le1/4+\int_2^{\infty}x^{-2}\,dx=3/4\). Replace each pair by the vector \((a,b)=(3u+2v,u+v)\). The matrix with rows \((3,2)\) and \((1,1)\) has determinant \(1\), so this vector is primitive and distinct pairs give distinct slopes. Its slope \((u+v)/(3u+2v)\) lies strictly between \(1/3\) and \(1/2\) because \(u,v\ge1\). Arrange these vectors by decreasing slope and concatenate them to form the edges of a concave polygonal function \(f:[0,L]\to\mathbb R\) with \(f(0)=0\). Its vertices \((x_i,y_i)\), \(0\le i\le r_m\), have integer coordinates, and
\begin{equation}\label{eq:polygonlength}
L=\sum a\le5m r_m\le5m^3.
\end{equation}
Define a binary word \(w=w_1\cdots w_L\) by \(w_j=\lfloor f(j)\rfloor-\lfloor f(j-1)\rfloor\). Each letter is \(0\) or \(1\), since every slope of \(f\) lies in \((1/3,1/2)\). Consider the deterministic graph whose unique infinite path has cost word \(11w0^\omega\); it uses \(L+3\) vertices.

For \(0\le\rho\le1/2\), put
\begin{equation}\label{eq:prefixsupport}
T(\rho)=\max_{0\le j\le L}\{\lfloor f(j)\rfloor-\rho j\}=\max_{0\le i\le r_m}\{y_i-\rho x_i\}.
\end{equation}
The equality holds because the integer points \((j,\lfloor f(j)\rfloor)\) lie on or below the polygon, and every polygon vertex occurs among these points. Concavity and \(f(0)=0\) give \(f(j)-f(i)\le f(j-i)\) for \(0\le i<j\le L\), so every window \(w_{i+1}\cdots w_j\) satisfies
\begin{equation}\label{eq:interiorwindow}
\sum_{t=i+1}^{j}w_t=\lfloor f(j)\rfloor-\lfloor f(i)\rfloor\le\lfloor f(j-i)\rfloor+1.
\end{equation}
Its excess is therefore at most \(\lfloor f(j-i)\rfloor-\rho(j-i)+1\le T(\rho)+1\). A prefix containing the two initial cost-\(1\) edges has maximum excess \(2(1-\rho)+T(\rho)\), which is at least \(T(\rho)+1\) on this rate interval. Any window meeting the initial block of ones can be extended to the beginning without decreasing its excess, and deleting trailing zeros never decreases excess. Hence the graph has boundary
\begin{equation}\label{eq:singlelassolower}
\Bstar(\rho)=2(1-\rho)+T(\rho)\qquad(0\le\rho\le1/2).
\end{equation}
The \(r_m\) distinct slopes of the polygon produce \(r_m\) distinct breakpoints of \(T\) in \((1/3,1/2)\), so \(\Bstar\) has at least \(r_m+1\) affine pieces. For an arbitrary sufficiently large \(N\), choose \(m=\lfloor((N-3)/5)^{1/3}\rfloor\) and append cost-\(0\) edges before the self-loop until the graph has exactly \(N\) vertices. These edges leave the burst requirement unchanged, and \eqref{eq:coprimecount} gives \(\Omega(N^{2/3})\) pieces.
\end{proof}

\section{The Farey structure of the burst boundary}\label{sec:frontier}

By Lemma~\ref{lem:lassocap}, every finite burst requirement is a maximum of lines from the family
\begin{equation}\label{eq:lines}
\mathcal A_N=\{k-\ell\rho:0\le k\le\ell\le N-1\},
\end{equation}
which contains the zero line \(k=\ell=0\). Two nonparallel lines of \(\mathcal A_N\) meet at \(\rho=(k-k')/(\ell-\ell')\), a fraction with denominator at most \(N-1\), and every cycle mean has denominator at most \(N\). All these rates in \([0,1]\) belong to the Farey sequence of order \(N\),
\begin{equation}\label{eq:farey}
\Farey_N=\left\{a/b:0\le a\le b\le N,\ \gcd(a,b)=1\right\},
\end{equation}
which has \(O(N^2)\) elements. We call the open intervals between consecutive elements of \(\Farey_N\) Farey intervals.

\begin{thm}\label{thm:frontier}
The burst boundary has the following properties.
\begin{enumerate}
\item It is nonincreasing and right-continuous on \([0,1)\), with values in \([0,\infty]\).
\item On each Farey interval, it is either identically infinite or equal to one line \(k-\ell\rho\) of \(\mathcal A_N\).
\item It consists of \(O(N^2)\) affine pieces, and a downward jump can occur only at the cycle mean of a simple lasso.
\item Every finite value satisfies
\begin{equation}\label{eq:sharpburst}
\Bstar(\rho)\le(N-1)(1-\rho).
\end{equation}
\end{enumerate}
\end{thm}

\begin{proof}
For (1), increasing \(\rho\) weakens \eqref{eq:budget}, so \(\Bstar\) is nonincreasing. Each \(\capreq_\lambda\) is infinite below \(\mu_\lambda\) and continuous on \([\mu_\lambda,1]\), hence right-continuous with values in \([0,\infty]\), and so is the minimum \eqref{eq:frontiermin} of finitely many such functions.

For (2), no cycle mean lies in a Farey interval, so the set of lassos with finite requirement is the same at every rate of the interval. No two lines of \(\mathcal A_N\) cross inside the interval, so each finite \(\capreq_\lambda\) equals one line of \(\mathcal A_N\) there, and so does their minimum.

For (3), \(\Farey_N\) has \(O(N^2)\) elements, so (2) gives \(O(N^2)\) pieces. Near a rate that is not a cycle mean, the set of lassos with finite requirement is constant, so \(\Bstar\) is a minimum of continuous functions there and does not jump.

For (4), each line in \eqref{eq:lassocap} satisfies \(k-\ell\rho\le\ell(1-\rho)\le(N-1)(1-\rho)\) because \(k\le\ell\le N-1\).
\end{proof}

The graph can have exponentially many simple lassos, but their pointwise minimum has only \(O(N^2)\) pieces. Each individual lasso has \(O(N^{2/3})\) pieces by Theorem~\ref{thm:singlelasso}; choosing among lassos adds pieces and can create discontinuities. The next corollary identifies the rate where the boundary becomes finite.

\begin{cor}\label{cor:mean}
Suppose \(G\) has a cycle, and let
\begin{equation}\label{eq:mu}
\mu=\min_{\beta}\frac{\cost(\beta)}{|\beta|}
\end{equation}
over simple cycles \(\beta\). Then \(\Bstar(\rho)<\infty\) if and only if \(\rho\ge\mu\). If \(G\) has no cycle, \(\Bstar\) is identically infinite.
\end{cor}

\begin{proof}
A lasso with finite requirement at \(\rho\) has cycle mean at most \(\rho\). Conversely, a simple path from \(S\) to a cycle attaining \(\mu\), stopped at its first vertex on that cycle, forms a simple lasso with finite requirement at every \(\rho\ge\mu\). An acyclic graph has no infinite path.
\end{proof}

Karp's algorithm computes \(\mu\) in \(O(NM)\) time \cite{karp1978}. By \eqref{eq:sharpburst}, a burst \(B\ge(N-1)(1-\rho)\) admits a feasible infinite path exactly when \(\rho\ge\mu\); smaller bursts require the value \(\Bstar(\rho)\).

\subsection{Jumps and the terminal piece}

\begin{prop}\label{prop:jumps}
For every \(\rho_0\in(0,1]\),
\begin{equation}\label{eq:leftlimit}
\Bstar(\rho_0^-)=\min_{\lambda:\,\mu_\lambda<\rho_0}\capreq_\lambda(\rho_0),
\end{equation}
where the minimum of an empty set is \(\infty\). If \(\Bstar(\rho_0)<\infty\), a downward jump, meaning \(\Bstar(\rho_0^-)>\Bstar(\rho_0)\), occurs at \(\rho_0\) exactly when every optimal lasso at \(\rho_0\) has cycle mean \(\rho_0\). At a reduced rational \(\rho_0=a/b\), every jump between two finite values has height a positive integer multiple of \(1/b\).
\end{prop}

\begin{proof}
The graph has finitely many simple lassos. Sufficiently close to \(\rho_0\) from the left, exactly those with \(\mu_\lambda<\rho_0\) have finite requirement, and each such requirement is continuous up to \(\rho_0\). Taking their minimum proves \eqref{eq:leftlimit}. Hence \(\Bstar(\rho_0^-)=\Bstar(\rho_0)\) exactly when some lasso with \(\mu_\lambda<\rho_0\) is optimal at \(\rho_0\). Every lasso with finite requirement at \(\rho_0\) has \(\mu_\lambda\le\rho_0\), so a jump occurs exactly when every optimal lasso has \(\mu_\lambda=\rho_0\). Finally, \eqref{eq:lassocap} evaluated at \(a/b\) is an integer multiple of \(1/b\) whenever finite, and the same holds for both minima.
\end{proof}

The ratio \(\Bstar(\rho)/(1-\rho)\) is the burst measured in units of a single erased step, since a block of \(k\) consecutive cost-\(1\) edges requires exactly \(k(1-\rho)\). The next result shows that this ratio is monotone and identifies its final value.

\begin{prop}\label{prop:terminal}
For \(0\le\rho<\sigma<1\) with \(\Bstar(\rho)<\infty\),
\begin{equation}\label{eq:normalized}
\frac{\Bstar(\sigma)}{1-\sigma}\le\frac{\Bstar(\rho)}{1-\rho}.
\end{equation}
Suppose an infinite path exists. Call the supremum of the lengths of the all-\(1\) windows of an infinite path its longest run, with value \(0\) if the path has no cost-\(1\) edge, and let \(R\) be the least longest run of an infinite path. If \(R<\infty\), then \(R\le N-1\), and
\begin{align}
\Bstar(\rho)&\ge R(1-\rho)&&(0\le\rho\le1),\label{eq:terminallower}\\
\Bstar(\rho)&=R(1-\rho)&&\Bigl(\frac{R}{R+1}\le\rho\le1\Bigr).\label{eq:terminal}
\end{align}
The threshold \(R/(R+1)\) is sharp. Once the boundary meets the line \(R(1-\rho)\), it follows the line at all larger rates.
\end{prop}

\begin{proof}
A window with \(k\) ones and \(z\) zeros has normalized excess \(k-z\rho/(1-\rho)\), which is nonincreasing in \(\rho\). Applying this to an optimal lasso at \(\rho\) proves \eqref{eq:normalized}. Let \(R<\infty\). Some cost-\(0\) edge occurs infinitely often on a path with bounded runs, so it lies on a cycle, and a simple lasso through a simple cycle containing it has runs of length at most \(N-1\); hence \(R\le N-1\). The integer minimum \(R\) is attained, and every infinite path contains an all-\(1\) window of length \(R\), which gives \eqref{eq:terminallower}. On a path whose longest run is \(R\), every window with \(k\) ones and \(z\) zeros satisfies \(k\le R(z+1)\), so
\begin{equation}\label{eq:runbound}
k-\rho(k+z)\le R(1-\rho)+z\bigl(R-(R+1)\rho\bigr).
\end{equation}
The last term is nonpositive for \(\rho\ge R/(R+1)\), which proves \eqref{eq:terminal}. For a graph whose only cycle has cost word \(1^R0\), the boundary is infinite below \(R/(R+1)\), so the threshold is sharp. The persistence of equality follows from \eqref{eq:normalized} and \eqref{eq:terminallower}.
\end{proof}

The two extreme cases are simple. If \(R=0\), an all-zero infinite path exists and \(\Bstar\) vanishes identically. If a cycle exists but \(R=\infty\), every cycle consists of cost-\(1\) edges, so \(\Bstar\) is infinite on \([0,1)\) and \(\Bstar(1)=0\).

\subsection{Shapes of the boundary}

Jumps are not the only source of nonconvexity. The next construction gives a boundary without jumps.

\begin{prop}\label{prop:continuous}
A seven-vertex graph has a boundary that is finite and continuous on \([0,1]\) and is neither convex nor concave.
\end{prop}

\begin{figure}[t]
\centering
\begin{tikzpicture}
\begin{axis}[width=7.5cm,height=5cm,xmin=0,xmax=1.03,ymin=0,ymax=3.15,xlabel={Rate \(\rho\)},ylabel={\(\Bstar(\rho)\)},xtick={0,0.3333333,0.5,1},xticklabels={\(0\),\(1/3\),\(1/2\),\(1\)},ytick={0,0.5,1.3333333,2,3},yticklabels={\(0\),\(1/2\),\(4/3\),\(2\),\(3\)},axis lines=left,tick label style={font=\footnotesize},label style={font=\small}]
\addplot[dashed,gray,domain=0:1]{2-2*x};
\addplot[dashed,gray,domain=0:0.5]{3-5*x};
\addplot[dashed,gray,domain=0.5:1]{1-x};
\addplot[very thick,teal!70!black] coordinates {(0,2)(0.3333333,1.3333333)(0.5,0.5)(1,0)};
\addplot[only marks,mark=*,mark size=1.8pt,teal!70!black] coordinates {(0.3333333,1.3333333)(0.5,0.5)};
\end{axis}
\end{tikzpicture}
\caption{The boundary of the two paths with cost words \(110^\omega\) and \(101010^\omega\). Dashed lines show the two individual requirements; their minimum is continuous, with slopes \(-2,-5,-1\).}
\label{fig:continuous}
\end{figure}
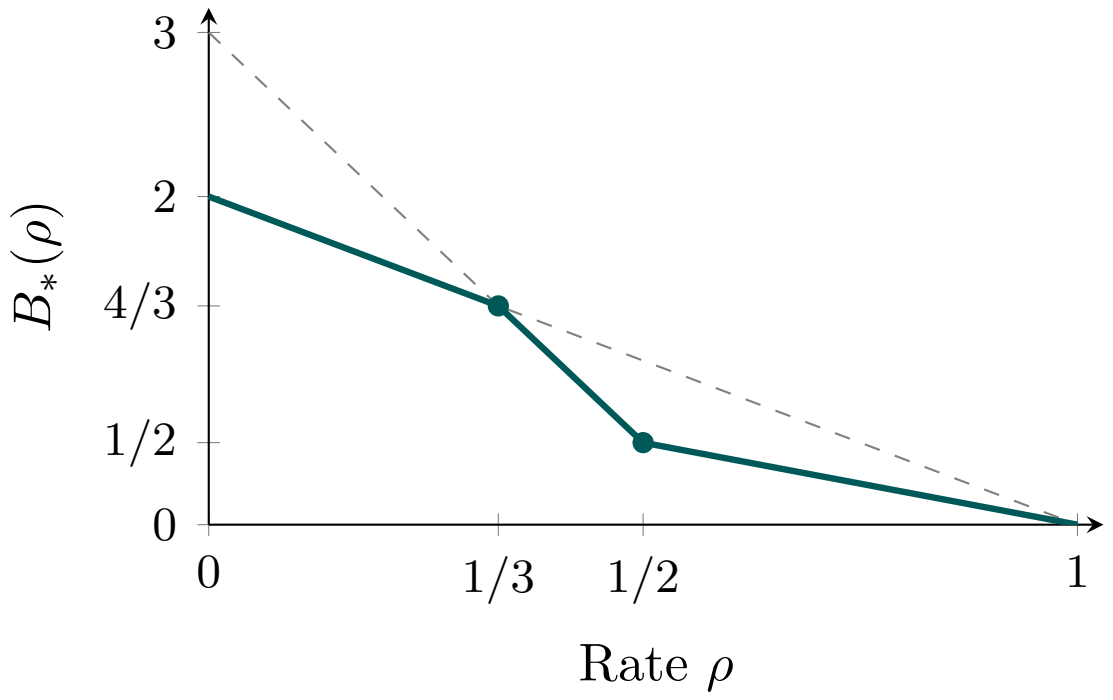

\begin{proof}
Take two internally disjoint paths from the initial vertex \(h\) to a vertex \(z\), with cost words \(11\) and \(10101\), and a cost-\(0\) self-loop at \(z\). The first path requires \(2-2\rho\). A positive-excess window of the second path can be trimmed to start and end with \(1\), so its requirement is \(\max\{1-\rho,2-3\rho,3-5\rho\}=\max\{1-\rho,3-5\rho\}\), because \(2-3\rho\) is the average of the other two lines. Taking the minimum gives
\begin{equation}\label{eq:continuous}
\Bstar(\rho)=\begin{cases}2-2\rho,&0\le\rho\le1/3,\\3-5\rho,&1/3\le\rho\le1/2,\\1-\rho,&1/2\le\rho\le1.\end{cases}
\end{equation}
The slopes \(-2,-5,-1\) prove both assertions about curvature; Figure~\ref{fig:continuous} plots the boundary.
\end{proof}

The next two constructions show that the bound \eqref{eq:sharpburst} and the Farey order \(N\) cannot be improved, and that the number of jumps can grow linearly in \(N\).

\begin{prop}\label{prop:sharp}
For every \(N\ge1\), some graph with \(N\) vertices satisfies \(\Bstar(\rho)=(N-1)(1-\rho)\) for all \(\rho\in[0,1]\), and some graph with \(N\) vertices has a breakpoint with denominator \(N\).
\end{prop}

\begin{proof}
A chain of \(N-1\) cost-\(1\) edges from the initial vertex to a cost-\(0\) self-loop has one infinite path, whose largest excess is attained by the chain itself. A directed \(N\)-cycle with cost word \(10^{N-1}\) has mean \(1/N\), where \(\Bstar\) changes from infinite to finite.
\end{proof}

\begin{prop}\label{prop:manypieces}
For every integer \(m\ge1\), let \(\mu_a=a/(3a-1)\) for \(1\le a\le m\). Some graph with \(N=3m-1\) vertices has the burst boundary
\begin{equation}\label{eq:manypieces}
\Bstar(\rho)=\begin{cases}\infty,&0\le\rho<\mu_m,\\a(1-\rho),&\mu_a\le\rho<\mu_{a-1}\quad(2\le a\le m),\\1-\rho,&\mu_1\le\rho\le1.\end{cases}
\end{equation}
Hence \(\Bstar\) has \(m=(N+1)/3\) finite pieces and \(m-1\) downward jumps.
\end{prop}

\begin{figure}[t]
\centering
\begin{tikzpicture}[>=Stealth,vertex/.style={circle,draw,minimum size=7mm,inner sep=1pt},every node/.style={font=\small}]
\node[vertex] (h) at (0,0) {\(h\)};
\node[vertex] (o1) at (1.8,1.4) {\(o_1\)};
\node[vertex] (o2) at (3.6,1.4) {\(o_2\)};
\node[vertex] (o3) at (5.4,1.4) {\(o_3\)};
\node[vertex] (y1) at (1.4,-1.4) {\(y_1\)};
\node[vertex] (y2) at (2.8,-1.4) {\(y_2\)};
\node[vertex] (y3) at (4.2,-1.4) {\(y_3\)};
\node[vertex] (y4) at (5.6,-1.4) {\(y_4\)};
\draw[->] (-0.9,0.8)--(h);
\draw[->] (h) to[bend left=18] node[above left]{\(1\)} (o1);
\draw[->] (o1) to[bend left=18] node[below right]{\(0\)} (h);
\draw[->] (o1)--node[above]{\(1\)}(o2);
\draw[->] (o2)--node[above]{\(1\)}(o3);
\draw[->] (o2)--node[right]{\(0\)}(y2);
\draw[->] (o3)--node[right]{\(0\)}(y4);
\draw[->] (y4)--node[below]{\(0\)}(y3);
\draw[->] (y3)--node[below]{\(0\)}(y2);
\draw[->] (y2)--node[below]{\(0\)}(y1);
\draw[->] (y1)--node[below left]{\(0\)}(h);
\end{tikzpicture}
\caption{The graph of Proposition~\ref{prop:manypieces} for \(m=3\). Its simple cycles have cost words \(10\), \(11000\), and \(11100000\), with mean costs \(1/2\), \(2/5\), and \(3/8\). The burst boundary is \(3-3\rho\) on \([3/8,2/5)\), \(2-2\rho\) on \([2/5,1/2)\), and \(1-\rho\) on \([1/2,1]\).}
\label{fig:manypieces}
\end{figure}
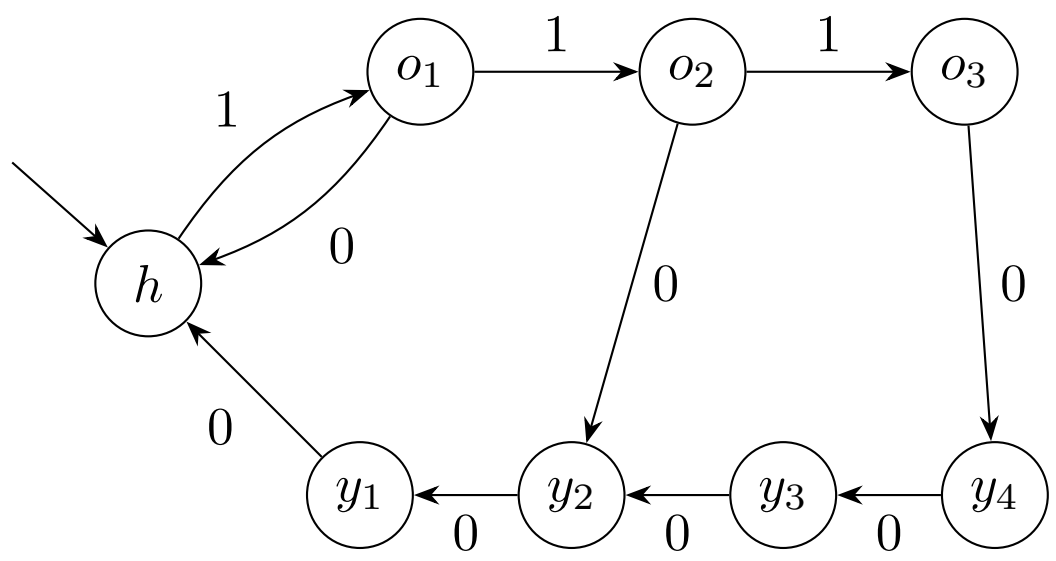

\begin{proof}
Take vertices \(h,o_1,\ldots,o_m,y_1,\ldots,y_{2m-2}\) with initial vertex \(h\), and write \(y_0=h\). The cost-\(1\) edges are \(h\to o_1\) and \(o_{a-1}\to o_a\) for \(2\le a\le m\). The cost-\(0\) edges are \(o_a\to y_{2a-2}\) for \(1\le a\le m\) and \(y_d\to y_{d-1}\) for \(1\le d\le2m-2\). Figure~\ref{fig:example} shows the case \(m=2\), and Figure~\ref{fig:manypieces} the case \(m=3\). Every cycle passes through \(h\), and the simple cycles are the paths that climb from \(h\) to some \(o_a\) and descend through \(y_{2a-2},\ldots,y_1\) to \(h\). The cycle through \(o_a\) has cost word \(\beta_a=1^a0^{2a-1}\), length \(3a-1\), and mean \(\mu_a\). Since \(h\) lies on every cycle, the simple lassos are the paths \(\beta_a^\omega\).

Fix \(a\) and \(\rho\ge\mu_a\). By the proof of Lemma~\ref{lem:lassocap}, it suffices to consider windows of \(\beta_a\beta_a\) with fewer than \(3a-1\) edges. Such a window that meets both blocks of ones contains the \(2a-1\) zeros between them and hence at most \(a-1\) ones. Every such window therefore has some number \(j\le a\) of ones, so its excess is at most \(j(1-\rho)\le a(1-\rho)\), and the block \(1^a\) attains this value. Thus \(\capreq_{\beta_a^\omega}(\rho)=a(1-\rho)\) for \(\rho\ge\mu_a\). The means \(\mu_a=1/(3-1/a)\) decrease strictly in \(a\), so at rate \(\rho\) the lassos with finite requirement are those with \(a\ge a_\rho\), where \(a_\rho\) is the least \(a\) with \(\mu_a\le\rho\). Corollary~\ref{cor:minlasso} gives \(\Bstar(\rho)=a_\rho(1-\rho)\), which is \eqref{eq:manypieces}.
\end{proof}

The jumps have a simple cause. At \(\rho=\mu_{a-1}\), the cycle \(\beta_{a-1}\) becomes usable. Its mean cost is larger than those of the cycles usable before, and its block of ones is shorter, so its burst requirement is smaller. The same mechanism produces the jump in Figure~\ref{fig:example}. A jump inside the finite domain rules out both convexity and concavity, since convex and concave functions are continuous on open intervals.

\subsection{The critical-rate function}

Fixing the burst instead of the rate leads to the least rate that admits a feasible infinite path. Assume that \(G\) has a cycle, and define
\begin{equation}\label{eq:defRstar}
\Rstar(B)=\min\{\rho\in[0,1]:B\ge\Bstar(\rho)\}\qquad(B\ge0).
\end{equation}
In the discrimination problem, \(\Rstar(B)\) is the least sustained erasure rate at which a channel with burst \(B\) can confuse the two modes forever. Theorem~\ref{thm:inverse} shows that the minimum exists and that \(\Rstar\) is continuous, although \(\Bstar\) can jump.

\begin{thm}\label{thm:inverse}
For each lasso \(\lambda\), let
\begin{equation}\label{eq:lassorate}
\rho_\lambda(B)=\max\Bigl(\{\mu_\lambda,0\}\cup\Bigl\{\frac{k-B}{\ell}:(k,\ell)\in\Intervals_\lambda\Bigr\}\Bigr).
\end{equation}
Then
\begin{equation}\label{eq:inverse}
\Rstar(B)=\min_{\lambda\in\Lassos}\rho_\lambda(B).
\end{equation}
The function \(\Rstar:[0,\infty)\to[0,1]\) is continuous, nonincreasing, and piecewise affine, and its slopes belong to \(\{0\}\cup\{-1/\ell:1\le\ell\le N-1\}\). Moreover, \(\Rstar(B)=\mu\) for every \(B\ge\Bstar(\mu)\), and a feasible infinite path exists at \(\rho=\Rstar(B)\).
\end{thm}

\begin{proof}
For a fixed lasso, \eqref{eq:lassocap} is at most \(B\) exactly when \(\rho\ge\mu_\lambda\) and \(\rho\ge(k-B)/\ell\) for each pair in \(\Intervals_\lambda\). Each of these lower bounds is at most \(1\), because \(k\le\ell\) and \(B\ge0\). Thus \(\rho_\lambda(B)\) is the least rate at which \(\lambda\) is \((\rho,B)\)-feasible. Taking the minimum over lassos proves \eqref{eq:inverse} and shows that the minimum in \eqref{eq:defRstar} is attained.

Each \(\rho_\lambda\) is a finite maximum of continuous affine functions of \(B\), and there are finitely many lassos. Their minimum is continuous and piecewise affine. On any open affine piece, the active expression is either a constant cycle mean or a line \((k-B)/\ell\), which gives the listed slopes. Increasing \(B\) weakens the constraint, so \(\Rstar\) is nonincreasing. Finally, no rate below \(\mu\) works by Corollary~\ref{cor:mean}, whereas \(\rho=\mu\) works as soon as \(B\ge\Bstar(\mu)\).
\end{proof}

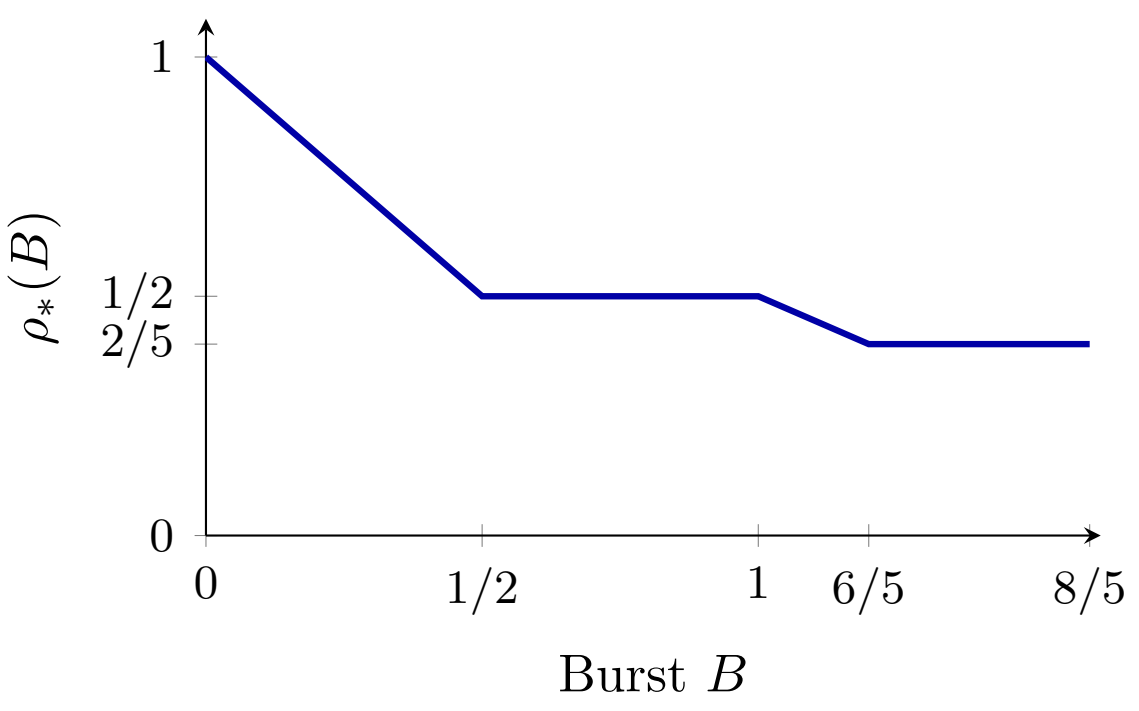
\begin{figure}[t]
\centering
\begin{tikzpicture}
\begin{axis}[width=7.5cm,height=5cm,xmin=0,xmax=1.62,ymin=0,ymax=1.08,xlabel={Burst \(B\)},ylabel={\(\Rstar(B)\)},xtick={0,0.5,1,1.2,1.6},xticklabels={\(0\),\(1/2\),\(1\),\(6/5\),\(8/5\)},ytick={0,0.4,0.5,1},yticklabels={\(0\),\(2/5\),\(1/2\),\(1\)},axis lines=left,tick label style={font=\footnotesize},label style={font=\small}]
\addplot[very thick,blue!65!black] coordinates {(0,1)(0.5,0.5)(1,0.5)(1.2,0.4)(1.6,0.4)};
\end{axis}
\end{tikzpicture}
\caption{The critical-rate function of the graph in Figure~\ref{fig:example}. The downward jump of \(\Bstar\) at \(\rho=1/2\) becomes the plateau on \([1/2,1]\).}
\label{fig:inverse}
\end{figure}

For the graph of Figure~\ref{fig:example}, the jump of \(\Bstar\) at \(1/2\) becomes a plateau of \(\Rstar\):
\begin{equation}\label{eq:inverseexample}
\Rstar(B)=\begin{cases}1-B,&0\le B\le1/2,\\1/2,&1/2\le B\le1,\\1-B/2,&1\le B\le6/5,\\2/5,&B\ge6/5.\end{cases}
\end{equation}
Figure~\ref{fig:inverse} plots this function. The set of feasible pairs \((\rho,B)\) is not convex: it contains \((2/5,6/5)\) and \((1/2,1/2)\), but not their midpoint \((9/20,17/20)\), since \(\Bstar(9/20)=11/10\).

\section{Computing the boundary}\label{sec:algorithm}

The algorithm evaluates \(\Bstar\) at one rational rate per Farey interval. At a rational rate \(\rho=p/q\) in lowest terms, the scaled load \(j_t=qx_t\) is an integer and satisfies
\begin{equation}\label{eq:integerload}
j_0=0,\qquad j_t=\max\{0,j_{t-1}+\delta_{e_t}\},\qquad\delta_e=qc(e)-p.
\end{equation}
For an integer capacity \(K\ge0\), which plays the role of the scaled burst \(qB\), the load graph \(G_{p,q,K}\) has states \((v,j)\) with \(v\in V\) and \(0\le j\le K\), and an edge from \((v,j)\) to \((u,\max\{0,j+\delta_e\})\) whenever \(e=(v,u)\in E\) and the new load is at most \(K\). Its initial states are \(S\times\{0\}\). Thus \(G_{p,q,K}\) is the product of \(G\) with the automaton \(D_{p,q,K}\) of Theorem~\ref{thm:regular}. Paths of \(G_{p,q,K}\) from initial states correspond to \((p/q,B)\)-feasible paths of \(G\) for every \(B\) with \(\lfloor qB\rfloor=K\), because the scaled loads are integers. A \((p/q,B)\)-feasible infinite path therefore exists exactly when \(G_{p,q,\lfloor qB\rfloor}\) has an infinite path from an initial state, that is, a reachable cycle. A graph search decides this in \(O((N+M)(K+1))\) time. Section~\ref{sec:thresholds} shows that the states from which an infinite path starts are described by one integer per vertex, which leads to a sweep that handles all capacities at once.

\subsection{Decreasing capacities with one threshold per vertex}\label{sec:thresholds}

Call a load state viable if an infinite path starts there. If \((v,j)\) is viable, every \((v,i)\) with \(0\le i\le j\) is viable, since every update in \eqref{eq:integerload} is nondecreasing in the load. Write \(\theta_v(K)\) for the largest viable load at \(v\), or \(-1\) if no load is viable, and call \(v\) live if \(\theta_v\ge0\) and dead otherwise. For a vector \(\theta\in\{-1,0,\ldots,K\}^{V}\), define
\begin{equation}\label{eq:thresholdmap}
A_v(\theta)=\max_{\substack{e=(v,u)\in E\\\theta_u\ge0}}(\theta_u-\delta_e),\qquad F_K(\theta)_v=\max\{-1,\min\{K,A_v(\theta)\}\},
\end{equation}
where a maximum over an empty set is \(-\infty\). An edge \(e=(v,u)\) supports the current threshold at \(v\) if
\begin{equation}\label{eq:support}
\theta_v\ge0,\qquad\theta_u\ge0,\qquad\theta_v\le\theta_u-\delta_e.
\end{equation}
These inequalities mean that every load at most \(\theta_v\) can use \(e\) and reach a load at most \(\theta_u\).

\begin{lem}\label{lem:thresholds}
The vector \(\theta(K)\) is the greatest fixed point of \(F_K\). Starting from any vector in \(\{-1,0,\ldots,K\}^V\) that bounds \(\theta(K)\) from above, repeatedly replacing an unsupported live component \(\theta_v\) by \(F_K(\theta)_v\) yields \(\theta(K)\).
\end{lem}

\begin{proof}
From load \(j\ge0\), an edge \(e=(v,u)\) to a live vertex \(u\) leads to a load in \([0,\theta_u]\) exactly when \(j\le\theta_u-\delta_e\). Hence \(F_K(\theta)_v\) is the largest load at \(v\) with an edge into the states retained by \(\theta\), and \(F_K\) is monotone. Every viable state has a viable successor, so \(\theta(K)\le F_K(\theta(K))\). If \(\theta\ge\theta(K)\), monotonicity gives \(F_K(\theta)\ge\theta(K)\), so no update removes a viable state. An unsupported live component satisfies \(F_K(\theta)_v<\theta_v\), so every update strictly lowers an integer component and the process terminates. At termination every live vertex has a supporting edge, which serves all loads up to its threshold. Following such edges gives an infinite path from every retained state, so the final vector equals \(\theta(K)\). The same argument shows that the states retained by any fixed point are viable, so \(\theta(K)\), which is itself a fixed point, is the greatest one.
\end{proof}

The states retained by \(\theta(K)\) form the greatest fixed point of the predecessor operator of the load graph, which is the standard characterization of the states from which a safety objective can be met. The threshold vector stores this fixed point with one integer per vertex instead of one bit per load state.

\begin{thm}\label{thm:capacitysweep}
Fix \(p/q\) and an integer \(K_{\max}\ge0\). The least feasible integer capacity in \([0,K_{\max}]\), or the fact that none exists, can be computed in
\begin{equation}\label{eq:sweeptime}
O((N+M)(K_{\max}+1))
\end{equation}
arithmetic operations and \(O(N+M)\) words of working storage. At the least feasible capacity, the algorithm also constructs an optimal simple lasso.
\end{thm}

\begin{proof}
Initialize \(\theta_v=K_{\max}\) for every vertex. For \(K=K_{\max},K_{\max}-1,\ldots,0\), first replace \(\theta_v\) by \(\min\{\theta_v,K\}\), leaving dead components \(-1\) unchanged. Then apply Lemma~\ref{lem:thresholds}. The resulting vector is \(\theta(K)\) by Lemma~\ref{lem:thresholds}: the initial vector bounds \(\theta(K_{\max})\) from above, and \(\theta(K)\le\min\{\theta(K+1),K\}\) because every state viable for \(K\) is viable for \(K+1\). Capacity \(K\) is feasible exactly when some \(s\in S\) has \(\theta_s(K)\ge0\). If the first stage is infeasible, report that no capacity in the range works. Otherwise save the last feasible vector; the first infeasible stage \(K\) identifies the optimum \(K+1\). If stage \(0\) remains feasible, the optimum is \(0\).

To implement a stage, store the support indicator \eqref{eq:support} for each edge and the number of supporting outgoing edges at each vertex. Rebuild these indicators after clamping, in \(O(N+M)\) time, and queue every live vertex with support count zero. When such a vertex \(v\) is removed from the queue, scan its outgoing edges to compute \(F_K(\theta)_v\). After lowering \(\theta_v\), update the indicators on all edges incident to \(v\) and the support counts of their sources. Queue each affected live vertex whose count is zero, including \(v\) itself if it remains unsupported. Ignore an extracted vertex if it has died or acquired support. Queue membership marks avoid duplicate entries.

Labels never increase during the sweep and take values in \(\{-1,\ldots,K_{\max}\}\), so each label decreases at most \(K_{\max}+1\) times. A decrease at \(v\) costs \(O(1+\deg^+(v)+\deg^-(v))\). Summing over vertices and decreases, and adding the \(K_{\max}+1\) stage initializations, gives \eqref{eq:sweeptime}. Queue operations are charged to these initializations and incident-edge updates. The labels, saved vector, adjacency lists, support indicators, counts, and queue occupy \(O(N+M)\) words.

At the optimum, choose one supporting outgoing edge at every live vertex in the saved vector. Start at a live initial vertex with load \(0\). The chosen edges preserve the threshold inequalities and hence feasibility. Since the choice depends only on the vertex, the resulting infinite path is a simple lasso. Its scaled requirement \(q\capreq_\lambda(p/q)\) is an integer by Lemma~\ref{lem:lassocap} and equals the least feasible capacity; a smaller value would contradict minimality.
\end{proof}

\subsection{Recovering an interval from its mediant}

\begin{lem}\label{lem:mediant}
Let \(a/b<c/d\) be consecutive elements of \(\Farey_N\), and let \(p/q=(a+c)/(b+d)\). Then \(p/q\) is in lowest terms, lies in \((a/b,c/d)\), and satisfies
\begin{equation}\label{eq:denominator}
N<q\le2N.
\end{equation}
If \(\Bstar\) equals \(k-\ell\rho\) on \((a/b,c/d)\), then \(K_*=q\Bstar(p/q)\) is an integer, and \(K_*\) determines \(k\) and \(\ell\) by
\begin{equation}\label{eq:decode}
\ell\equiv-p^{-1}K_*\pmod q,\quad0\le\ell<q,\qquad k=(K_*+p\ell)/q.
\end{equation}
\end{lem}

\begin{proof}
Consecutive Farey fractions satisfy \(bc-ad=1\) and \(b+d>N\) \cite[Theorems~28 and~30]{hardy2008}. Every common divisor of \(a+c\) and \(b+d\) divides \(bc-ad=1\), so the mediant is reduced. It lies strictly between \(a/b\) and \(c/d\), and \(b,d\le N\) gives \eqref{eq:denominator}. By Theorem~\ref{thm:frontier}, \(K_*=qk-p\ell\), so \(K_*\equiv-p\ell\pmod q\). Since \(p\ge1\) is invertible modulo \(q\) and \(0\le\ell\le N-1<q\), the congruence determines \(\ell\), and then \(k\) follows.
\end{proof}

For the graph of Figure~\ref{fig:example}, \(N=5\), and \(2/5\) and \(1/2\) are consecutive in \(\Farey_5\). Their mediant is \(3/7\), where \(\delta_e=4\) on cost-\(1\) edges and \(\delta_e=-3\) on cost-\(0\) edges. At \(K=8\), the thresholds at \(h,o_1,o_2,y_2,y_1\) are \(0,4,8,6,3\), and each vertex is supported by its edge on the cycle \(11000\). At \(K=7\), clamping \(\theta_{o_2}\) to \(7\) leaves \(o_1\) unsupported, its threshold drops to \(3\), and then \(\theta_h\) drops to \(-1\); the vertices \(y_1,y_2,o_2,o_1\) die in this order. Hence the sweep returns \(K_*=7\Bstar(3/7)=8\). Since \(3^{-1}\equiv5\pmod7\), formula \eqref{eq:decode} gives \(\ell\equiv-40\equiv2\pmod7\) and \(k=(8+3\cdot2)/7=2\), which recovers the line \(2-2\rho\) on \((2/5,1/2)\).

\subsection{The complete algorithm}

\begin{algo}[Burst boundary]\label{alg:boundary}\hfill
\begin{enumerate}
\item Merge edges with the same endpoints and the same cost, and stream \(\Farey_N\) in increasing order.
\item For each consecutive pair \(a/b<c/d\), set \(p=a+c\), \(q=b+d\), and \(K_{\max}=(N-1)(q-p)\).
\item Run the sweep of Theorem~\ref{thm:capacitysweep} from \(K_{\max}\) downward. If \(K_{\max}\) is infeasible, record \(\infty\) on \((a/b,c/d)\).
\item Otherwise let \(K_*\) be the least feasible capacity, recover \(\ell\) and \(k\) by \eqref{eq:decode}, and record the line \(k-\ell\rho\) and the optimal lasso on \((a/b,c/d)\).
\item Give each grid point below \(1\) the value of the line on its right, set \(\Bstar(1)=0\) if \(G\) has a cycle and \(\Bstar(1)=\infty\) otherwise, and merge consecutive pieces with the same coefficients and compatible endpoint values.
\end{enumerate}
\end{algo}

\begin{thm}\label{thm:algorithm}
Algorithm~\ref{alg:boundary} computes the burst boundary using
\begin{equation}\label{eq:time}
O((N+M)N^4)
\end{equation}
arithmetic operations, \(O(N+M)\) words of working storage, and \(O(N^2)\) words for the output. After duplicate edges are merged, all integers have \(O(\log(N+1))\) bits. For each Farey interval on which the boundary is finite, the algorithm also yields a simple lasso that is optimal on the interval and at its left endpoint.
\end{thm}

\begin{proof}
Merging uses adjacency lists and an array indexed by the target and the cost, in \(O(N+M)\) time, and leaves at most \(2N^2\) edges. Every finite value of \(q\Bstar(p/q)\) is an integer by Lemma~\ref{lem:mediant} and is at most \(K_{\max}\) by \eqref{eq:sharpburst}. If no capacity in \([0,K_{\max}]\) is feasible, the boundary is infinite throughout the interval by Theorem~\ref{thm:frontier}. Otherwise the least capacity is \(K_*=q\Bstar(p/q)\), and \eqref{eq:decode} recovers \(k\) and \(\ell\).

Let \(\lambda\) be the optimal lasso returned at the mediant. The lassos with finite requirement are the same throughout the open Farey interval, and no distinct lines of \(\mathcal A_N\) cross there. The maximizing line of \(\lambda\) and the line defining \(\Bstar\) agree at the mediant, so Lemma~\ref{lem:mediant} implies that they have the same coefficients. Thus \(\lambda\) is optimal throughout the interval. Right continuity extends this equality to \(a/b\). At \(c/d\), the value \(\capreq_\lambda(c/d)=\Bstar(c/d^-)\) can exceed \(\Bstar(c/d)\).

The Farey sequence has \(O(N^2)\) intervals, and \(q\le2N\) gives \(K_{\max}=O(N^2)\). Theorem~\ref{thm:capacitysweep} therefore gives \eqref{eq:time}. To stream the Farey sequence, start at \(0/1,1/N\); from consecutive \(a/b,c/d\), the next fraction is \((tc-a)/(td-b)\) with \(t=\lfloor(N+b)/d\rfloor\), until \(1/1\) is reached. This rule follows from the mediant property of three consecutive terms \cite[Theorem~29]{hardy2008} and uses constant additional storage. The value at grid points below \(1\) follows from right continuity, and at \(\rho=1\) every infinite path is feasible with \(B=0\). Lassos may be returned as intervals are processed; storing all of them would require \(O(N^3)\) additional words. All loads, coefficients, denominators, and support counts are polynomially bounded in \(N\).
\end{proof}

\begin{cor}\label{cor:query}
After \(O(N^2)\) preprocessing to store \(\Farey_N\), the value of \(\Bstar\) at an arbitrary rational rate \(\rho\in[0,1]\) can be obtained with \(O(\log(N+1))\) comparisons against that grid, \(O((N+M)N^2)\) arithmetic operations on integers of \(O(\log(N+1))\) bits, and one evaluation of an affine function at the input rational. Every load graph used has \(O(N^3)\) states, whatever the denominator of \(\rho\).
\end{cor}

\begin{proof}
For \(\rho<1\), locate \(a/b\le\rho<c/d\) by binary search in the grid. Compute the boundary only at the mediant of that interval. Return \(\infty\) if that value is infinite; otherwise decode its line and evaluate it at \(\rho\). Right continuity covers the case \(\rho=a/b\). At \(\rho=1\), only a cycle test is needed. The comparisons and the final evaluation use the bit length of the input; the sweep depends only on \(N\) and \(M\).
\end{proof}

Comparing the returned value with a rational burst \(B\) decides uniform distinguishability at \((\rho,B)\) by Proposition~\ref{prop:compact}. A direct test at \(\rho=p/q\) would build \(G_{p,q,\lfloor qB\rfloor}\), whose size grows with \(q\); the corollary avoids this growth because it evaluates \(\Bstar\) only at a mediant with denominator at most \(2N\).

\begin{rem}\label{rem:weights}
For integer costs in \(\{0,\ldots,W\}\), with \(W\ge1\), and rates in \([0,W]\), Theorem~\ref{thm:lasso} and Lemma~\ref{lem:lassocap} hold unchanged, and the lines satisfy \(0\le k\le W\ell\) and \(\ell\le N-1\). The breakpoints lie in the \(W\) integer translates of \(\Farey_N\), consecutive grid points still satisfy the determinant identity, and \(K_{\max}=(N-1)(Wq-p)\). The sweep gives \(O((N+M)W^2N^4)\) arithmetic operations and \(O(N+M)\) working words, with \(O(WN^2)\) output words. This dependence is polynomial in \(W\), rather than in its bit length. In the discrimination problem the costs are binary by \eqref{eq:binarycost}, so \(W=1\).
\end{rem}

\section{Discrimination horizons}\label{sec:delay}

When no feasible infinite path exists, the horizon \(H(\rho,B)\) of Definition~\ref{def:horizon} measures how long ambiguity can last. This section computes \(H\) exactly at rational parameters and determines its growth as \(B\) increases.

\begin{prop}\label{prop:horizon}
Let \(\rho=p/q\) and \(B\ge0\) be rational, and set \(K=\lfloor qB\rfloor\). If the part of \(G_{p,q,K}\) reachable from its initial states contains a cycle, then \(H(\rho,B)=\infty\). Otherwise, let \(L\) be the length of a longest path from an initial state. Then
\begin{equation}\label{eq:exacthorizon}
H(\rho,B)=L+1\le N(K+1),
\end{equation}
and \(H\) can be computed in \(O((N+M)(K+1))\) arithmetic operations.
\end{prop}

\begin{proof}
The cyclic case follows from Proposition~\ref{prop:compact} and the correspondence between paths of \(G_{p,q,K}\) and feasible paths of the comparison graph. In the acyclic case, dynamic programming in topological order computes the longest path, and Proposition~\ref{prop:compact} gives \(H=L+1\). A path in the acyclic reachable part visits at most \(N(K+1)\) states, so \(L\le N(K+1)-1\).
\end{proof}

The running time in Proposition~\ref{prop:horizon} is pseudopolynomial: it grows linearly with \(K\), and hence with \(q\). Theorem~\ref{thm:algorithm} avoids this growth because it chooses the denominators at which it evaluates \(\Bstar\). The next theorem gives the growth of \(H\) in closed form.

\begin{thm}\label{thm:delay}
Suppose the comparison graph has a cycle, let \(\mu\) be its minimum cycle mean \eqref{eq:mu}, fix \(0\le\rho<\mu\), and put
\begin{equation}\label{eq:delayconstants}
\delta=\mu-\rho>0,\qquad C=(N-1)(1-\mu).
\end{equation}
For every \(B\ge C\),
\begin{equation}\label{eq:delaybounds}
\Bigl\lfloor\frac{B-C}{\delta}\Bigr\rfloor+1\le H(\rho,B)\le\Bigl\lfloor\frac{B+\mu(N-1)}{\delta}\Bigr\rfloor+1,
\end{equation}
and the upper bound holds for every \(B\ge0\). In particular,
\begin{equation}\label{eq:asymptotic}
\lim_{B\to\infty}\frac{H(\rho,B)}{B}=\frac{1}{\mu-\rho}.
\end{equation}
\end{thm}

\begin{proof}
Take any feasible path of length \(n\). Remove cycles until a simple path remains. Its length is at most \(N-1\), and its cost is nonnegative. Every removed closed walk decomposes into simple cycles whose mean costs are at least \(\mu\). Therefore the original path has cost at least \(\mu(n-(N-1))\). Applying the window constraint to the whole path yields
\begin{equation}\label{eq:delayupper}
\mu(n-(N-1))\le\rho n+B,
\end{equation}
so \(n\le(B+\mu(N-1))/\delta\). This proves the upper bound by \eqref{eq:horizonL}.

For the lower bound, choose a simple cycle of mean \(\mu\) and a simple path from \(S\) to its first vertex on the cycle. They form a simple lasso \(\lambda\). At rate \(\mu\), Lemma~\ref{lem:lassocap} and \eqref{eq:sharpburst} bound the excess of every window of \(\lambda\) by \(C\). At rate \(\rho=\mu-\delta\), a window of length \(\ell\le n\) has excess at most \(C+\delta\ell\le C+\delta n\). Consequently, the first \(n=\lfloor(B-C)/\delta\rfloor\) edges of the lasso form a feasible path when \(B\ge C\), and \(H\ge n+1\). Dividing both bounds by \(B\) and letting \(B\) increase proves \eqref{eq:asymptotic}.
\end{proof}

For a single cost-\(1\) self-loop, \(N=1\) and \(\mu=1\), and every path of length \(n\) requires the burst \(n(1-\rho)\). Thus \(H(\rho,B)=\lfloor B/(1-\rho)\rfloor+1\) for \(0\le\rho<1\), and both bounds in \eqref{eq:delaybounds} are equalities. If the comparison graph is acyclic, ambiguity lasts at most \(N-1\) steps for all parameters, so \(H\le N\).

At a fixed burst, the horizon can instead jump from a finite value to infinity as the rate grows. Let the comparison graph be the two-edge cycle with cost word \(10\), with its cost-\(1\) edge first, and fix \(B=1/2\). If \(\rho<1/2\), the first difference cannot be erased, because the one-step constraint would require \(1\le\rho+1/2\). At \(\rho=1/2\), the reflected loads alternate between \(1/2\) and \(0\), so ambiguity persists forever. Hence
\begin{equation}\label{eq:horizonjump}
H(\rho,1/2)=\begin{cases}1,&0\le\rho<1/2,\\\infty,&1/2\le\rho\le1.\end{cases}
\end{equation}
Theorem~\ref{thm:delay} describes the growth of \(H\) along the other axis, as \(B\) increases at a fixed rate below \(\mu\).

\section{Rate--burst synthesis in safety games}\label{sec:games}

The Farey structure persists when a controller chooses some transitions and an adversary chooses the others. Consider a finite turn-based arena
\begin{equation}\label{eq:arena}
\mathcal G=(V_C,V_E,E,s,c),\qquad V=V_C\mathbin{\dot\cup}V_E,
\end{equation}
with initial vertex \(s\), edge costs \(c:E\to\{0,1\}\), and at least one outgoing edge at every vertex. Write \(N=|V|\) and \(M=|E|\). The controller selects an outgoing edge at vertices in \(V_C\), and the environment selects one at vertices in \(V_E\). Both players observe the current vertex and the executed edges. A strategy of the controller is winning at \((\rho,B)\) if every resulting infinite play satisfies \eqref{eq:budget}. Let
\begin{equation}\label{eq:gameboundary}
B_{\mathcal G}(\rho)=\inf\{B\ge0:\text{the controller has a winning strategy at }(\rho,B)\}.
\end{equation}
A cost-\(1\) edge marks a step that consumes one unit of a regulated resource, such as a message on a shaped link, and the game asks for the least burst under which the controller can keep the resource use of every run within the rate--burst bound. The controller observes the arena vertex, so the game models resource-constrained scheduling rather than diagnosis. After deleting the vertices from which no infinite path starts, the problem of Section~\ref{sec:boundary} with one initial vertex is the special case in which the controller owns every vertex.

\subsection{Positional strategies for real parameters}

A positional strategy selects one outgoing edge at each controller vertex, independently of the history and of the current load. The following lemma extends the proof of Theorem~\ref{thm:lasso} to two players.

\begin{lem}\label{lem:gamepositional}
For every real \(\rho\in[0,1]\) and \(B\ge0\), a winning strategy exists if and only if a winning positional strategy exists. A single positional strategy can be chosen to win from every configuration from which some strategy wins.
\end{lem}

\begin{proof}
Use the energy update \(T_e\) of \eqref{eq:energy}. Let \(W_v^h\subseteq[0,B]\) be the set of initial energies from which the controller can keep the energy nonnegative for \(h\) steps, starting at \(v\). Set \(W_v^0=[0,B]\). At a controller vertex, \(W_v^{h+1}\) is the union, over outgoing edges \(e=(v,u)\), of the sets \([0,B]\cap T_e^{-1}(W_u^h)\); at an environment vertex it is their intersection. Each \(T_e\) is continuous and nondecreasing, so every \(W_v^h\) is closed and upward closed, and these sets decrease with \(h\).

Put \(W_v=\bigcap_{h\ge0}W_v^h\). Every nonempty \(W_v\) has a minimum \(\tau_v\). If \(v\in V_C\), then for each \(h\) some outgoing edge \(e=(v,u)\) satisfies \(T_e(\tau_v)\in W_u^h\). Since the edge set is finite, one edge has this property for infinitely many \(h\), and nestedness gives \(T_e(\tau_v)\in W_u\), that is, \(T_e(\tau_v)\ge\tau_u\). Fix such an edge at \(v\). If \(v\in V_E\), every outgoing edge satisfies the same inequality. Monotonicity of \(T_e\) shows that these choices preserve \(E_t\ge\tau_{v_t}\) along every play that starts with \(E_0\in W_{v_0}\), so they win forever. Conversely, every configuration from which some strategy wins belongs to every finite-horizon set \(W_v^h\).
\end{proof}

\subsection{A finite expression for each positional strategy}

Fix a positional strategy \(\sigma\). Keep its chosen edge at each controller vertex, keep every environment edge, and restrict to the vertices reachable from \(s\). Denote the resulting graph by \(G_\sigma\). Let \(\mathcal P_\sigma\) be the set of nonempty simple paths of \(G_\sigma\), starting at any vertex of \(G_\sigma\), and define
\begin{equation}\label{eq:policymean}
\nu_\sigma=\max_{\beta\text{ a simple cycle of }G_\sigma}\frac{\cost(\beta)}{|\beta|}.
\end{equation}
The maximum exists because \(G_\sigma\) is finite and every vertex has an outgoing edge. A simple path has at most \(N-1\) edges.

\begin{lem}\label{lem:policycapacity}
The least burst for which \(\sigma\) is winning equals
\begin{equation}\label{eq:policycapacity}
F_\sigma(\rho)=\begin{cases}\infty,&\rho<\nu_\sigma,\\\max\bigl(\{0\}\cup\{\cost(P)-\rho|P|:P\in\mathcal P_\sigma\}\bigr),&\rho\ge\nu_\sigma.\end{cases}
\end{equation}
\end{lem}

\begin{proof}
If a cycle of \(G_\sigma\) has mean cost greater than \(\rho\), the environment can reach it and repeat it, and no finite burst bounds the resulting excess. Every path in \(\mathcal P_\sigma\) is a window of some play from \(s\), since its start vertex is reachable and every vertex has a successor, so its excess is a lower bound.

Conversely, let \(\rho\ge\nu_\sigma\) and take any window of any play in \(G_\sigma\). Remove a simple cycle whenever a vertex repeats. Every removed cycle has nonpositive excess, so removal does not decrease the excess of the window. The remaining walk is a simple path or is empty, so its excess is at most the maximum in \eqref{eq:policycapacity}.
\end{proof}

\begin{thm}\label{thm:gamefrontier}
For a turn-based safety arena with 0/1 costs,
\begin{equation}\label{eq:gamemin}
B_{\mathcal G}(\rho)=\min_{\sigma\text{ positional}}F_\sigma(\rho),
\end{equation}
and every finite value is attained. This boundary is nonincreasing and right-continuous, has \(O(N^2)\) affine pieces, satisfies \(B_{\mathcal G}(\rho)\le(N-1)(1-\rho)\) whenever it is finite, and on each Farey interval of order \(N\) is either infinite or one line of \(\mathcal A_N\). Moreover, each Farey interval on which \(B_{\mathcal G}\) is finite admits one positional strategy \(\sigma\) with \(F_\sigma=B_{\mathcal G}\) throughout the interval and at its left endpoint.
\end{thm}

\begin{proof}
There are finitely many positional strategies. By Lemmas~\ref{lem:gamepositional} and~\ref{lem:policycapacity}, the strategy \(\sigma\) wins exactly when \(B\ge F_\sigma(\rho)\), which proves \eqref{eq:gamemin} and attainment. Each threshold \(\nu_\sigma\) is the mean of a simple cycle and hence belongs to \(\Farey_N\), and every affine term in \eqref{eq:policycapacity} belongs to \(\mathcal A_N\). The proof of Theorem~\ref{thm:frontier} then applies verbatim.

Choose a strategy minimizing \(F_\sigma\) at an interior point of a Farey interval with finite boundary. No two distinct lines of \(\mathcal A_N\) cross inside the interval, and no threshold \(\nu_\sigma\) lies inside it, so the same strategy minimizes throughout. Its threshold is at most the left endpoint, and both \(F_\sigma\) and \(B_{\mathcal G}\) are right-continuous there, so equality also holds at the left endpoint.
\end{proof}

Let \(\nu_*=\min_\sigma\nu_\sigma\). Then \(B_{\mathcal G}(\rho)\) is finite exactly when \(\rho\ge\nu_*\). The chain and cycle of Proposition~\ref{prop:sharp}, with forced moves only, show that the burst bound and the order of the Farey grid are sharp for games as well.

\begin{thm}\label{thm:gamesynthesis}
The boundary \(B_{\mathcal G}\) and one optimal positional strategy for each Farey interval on which it is finite can be computed with
\begin{equation}\label{eq:gametime}
O((N+M)N^4\log(N+1))
\end{equation}
arithmetic operations on integers of \(O(\log(N+1))\) bits. The boundary uses \(O(N^2)\) output words, and storing all strategies explicitly uses \(O(N^3)\) words.
\end{thm}

\begin{proof}
At a mediant \(p/q\) and integer capacity \(K\), form the load game on states \((v,j)\) with \(0\le j\le K\). Each state keeps the owner of \(v\), transitions follow \eqref{eq:integerload}, and every move whose new load exceeds \(K\) leads to a losing absorbing state. Winning at \((p/q,B)\) with \(\lfloor qB\rfloor=K\) is exactly avoiding this state. The winning region of this safety game is computed by backward propagation: an environment state is losing as soon as one successor is losing, and a controller state is losing as soon as all its successors are losing. Maintaining successor counts takes time linear in the size of the load game, namely \(O((N+M)(K+1))\).

Theorem~\ref{thm:gamefrontier} gives \(K_{\max}=(N-1)(q-p)=O(N^2)\). Test this capacity and, if it is winning, binary-search the least winning integer \(K_*\); winning is monotone in \(K\). The least burst has the form \(k-\ell p/q\), so \(K_*=qk-p\ell\) is an integer, and the modular calculation of Lemma~\ref{lem:mediant} recovers the line. Repeating over all Farey intervals gives \eqref{eq:gametime}.

To extract a strategy, solve the load game once more at \(K_*\). For each vertex \(v\) with some winning load, the winning loads form an initial segment \(\{0,\ldots,h_v\}\), because a smaller load can only help. At a controller vertex, choose an edge that leads from \((v,h_v)\) to a winning state. By monotonicity of the load update, the same edge leads to a winning state from every \((v,j)\) with \(j\le h_v\). At an environment vertex, every successor of \((v,h_v)\) is winning, and the same holds for all smaller loads. The resulting positional strategy wins from \((s,0)\) and therefore attains the least burst at the mediant. By Theorem~\ref{thm:gamefrontier}, it attains the boundary throughout the interval and at its left endpoint.
\end{proof}

\subsection{An optimal strategy that changes between Farey intervals}

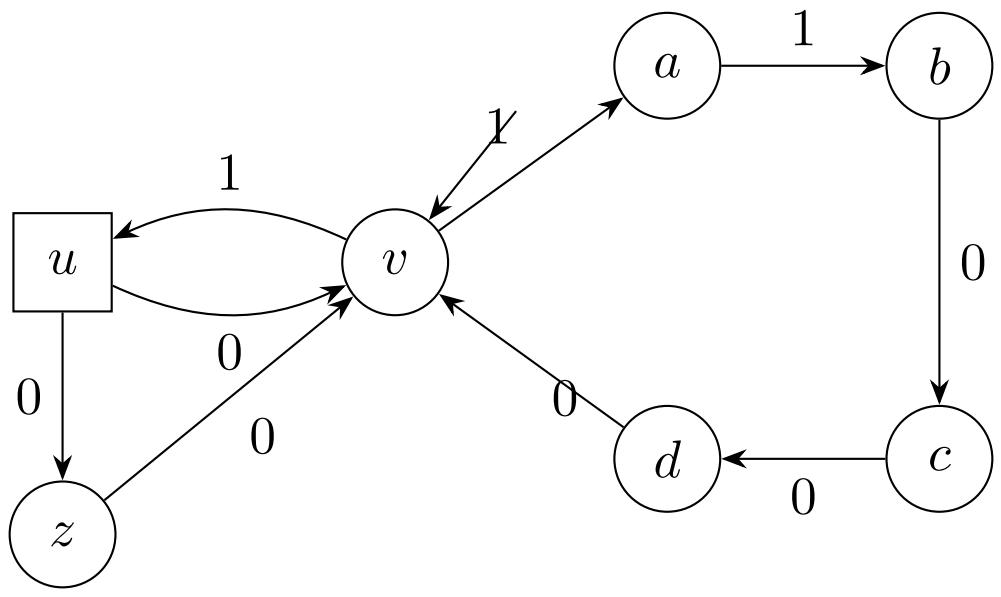
\begin{figure}[t]
\centering
\begin{tikzpicture}[>=Stealth,vertex/.style={circle,draw,minimum size=7mm,inner sep=1pt},env/.style={rectangle,draw,minimum size=6.5mm,inner sep=1pt},every node/.style={font=\small}]
\node[vertex] (v) at (0,0) {\(v\)};
\node[vertex] (a) at (1.8,1.3) {\(a\)};
\node[vertex] (b) at (3.6,1.3) {\(b\)};
\node[vertex] (c) at (3.6,-1.3) {\(c\)};
\node[vertex] (d) at (1.8,-1.3) {\(d\)};
\node[env] (u) at (-2.2,0) {\(u\)};
\node[vertex] (z) at (-2.2,-1.8) {\(z\)};
\draw[->] (0.8,1.0)--(v);
\draw[->] (v)--node[above left]{\(1\)}(a);
\draw[->] (a)--node[above]{\(1\)}(b);
\draw[->] (b)--node[right]{\(0\)}(c);
\draw[->] (c)--node[below]{\(0\)}(d);
\draw[->] (d)--node[below right]{\(0\)}(v);
\draw[->] (v) to[bend right=25] node[above]{\(1\)} (u);
\draw[->] (u) to[bend right=25] node[below]{\(0\)} (v);
\draw[->] (u)--node[left]{\(0\)}(z);
\draw[->] (z)--node[below right]{\(0\)}(v);
\end{tikzpicture}
\caption{A safety arena with controller vertex \(v\) and environment vertex \(u\) (square). The other vertices have forced moves. The optimal positional strategy takes the long route on \([2/5,1/2)\) and the short route on \([1/2,1)\).}
\label{fig:arena}
\end{figure}

The following seven-vertex arena, shown in Figure~\ref{fig:arena}, has the same boundary as the graph of Figure~\ref{fig:example}, but the environment has a real choice. The controller owns \(v\) and chooses between a long route \(v\to a\to b\to c\to d\to v\) with cost word \(11000\) and a short route that starts with a cost-\(1\) edge \(v\to u\). The environment owns \(u\) and either returns to \(v\) at cost \(0\) or moves at cost \(0\) to a vertex \(z\), which has a forced cost-\(0\) edge to \(v\). All other vertices have forced moves. On the short route, the environment thus chooses between the cost words \(10\) and \(100\).

The strategy that always takes the long route has threshold \(2/5\) and requirement \(2-2\rho\) above it. The strategy that always takes the short route has threshold \(1/2\), because below it the environment can repeat \(10\) forever. For \(\rho\ge1/2\), consecutive cost-\(1\) edges on the short route are separated by at least one cost-\(0\) edge, so a window with \(k\ge1\) ones has length at least \(2k-1\) and excess at most
\begin{equation}\label{eq:shortroute}
k-(2k-1)\rho=(1-\rho)+(k-1)(1-2\rho)\le1-\rho,
\end{equation}
and a single cost-\(1\) edge attains this bound. These are the only two positional strategies, so Theorem~\ref{thm:gamefrontier} gives the boundary of Figure~\ref{fig:example}. The long route is optimal on \([2/5,1/2)\), the short route is strictly better on \([1/2,1)\), and the optimal strategy changes at the Farey point \(1/2\).

\subsection{Büchi acceptance}

A Büchi condition requires a designated edge to occur infinitely often. Dziadek et al.~\cite{dziadek2025} study energy problems with such conditions. The following one-vertex example shows that a Büchi condition breaks right continuity and the bounded periods of Theorem~\ref{thm:lasso}.

\begin{prop}\label{prop:buchi}
Consider one vertex with a cost-\(0\) self-loop and a cost-\(1\) self-loop, and require the cost-\(1\) loop to occur infinitely often. The least burst is
\begin{equation}\label{eq:buchiboundary}
B_{\mathrm{acc}}(\rho)=\begin{cases}\infty,&\rho=0,\\1-\rho,&0<\rho\le1.\end{cases}
\end{equation}
This function is not right-continuous at \(0\), and at \(B=1\) the feasible rates have infimum \(0\) but no minimum. For \(\rho=1/q\) and \(B=1-1/q\), every accepted ultimately periodic cost sequence has a period of length at least \(q\).
\end{prop}

\begin{proof}
At \(\rho=0\), the prefix constraints bound the total number of cost-\(1\) edges by \(B\), so no accepted infinite play is feasible for finite \(B\). At \(\rho>0\), an accepted play contains a cost-\(1\) edge, whose one-step window requires \(B\ge1-\rho\). Let \(L=\lceil1/\rho\rceil\). The periodic word \((10^{L-1})^\omega\) is accepted and its period has excess \(1-L\rho\le0\). Removing complete periods from any window leaves a window of length less than \(L\) with at most one cost-\(1\) edge, whose excess is at most \(1-\rho\). This proves \eqref{eq:buchiboundary} and the claims about right continuity and the infimum.

Finally, an accepted ultimately periodic word has at least one cost-\(1\) edge in its repeated block. If the block has length \(d\) and contains \(k\ge1\) such edges, feasibility of arbitrarily many repetitions requires \(k/d\le1/q\). Hence \(d\ge qk\ge q\), although the arena has only one vertex.
\end{proof}

\section{Conclusion}\label{sec:conclusion}

The paper asked for the least burst at which a graph with 0/1 edge costs admits a feasible infinite path, as a function of the real rate. The feasible words form a safety language that is \(\omega\)-regular at rational rates and not at irrational rates once \(B\ge1\), so the lasso witnesses at irrational rates required a direct argument. The answer combines the convex window geometry of individual paths with discrete choices between cycles. A fixed simple lasso has a tight \(\Theta(N^{2/3})\) piece bound, while the choice among lassos can destroy both convexity and concavity and create downward jumps. All breakpoints lie in the Farey sequence of order \(N\), and one evaluation at the mediant of each Farey interval recovers the affine piece and an optimal lasso. A sweep over decreasing integer capacities, with one load threshold per vertex, computes the whole boundary in \(O((N+M)N^4)\) arithmetic operations and linear working space. For a monitor that distinguishes two finite-state modes through a leaky-bucket erasure channel, this boundary separates finite-time discrimination from permanent ambiguity, and below the minimum cycle mean the discrimination horizon grows as \(B/(\mu-\rho)\). The same Farey partition holds for turn-based safety games, with one optimal positional strategy per Farey interval, whereas a Büchi condition destroys right continuity.

Two questions remain open. The first is the maximum number of pieces of the boundary of an \(N\)-vertex graph, which lies between \(\Omega(N)\) and \(O(N^2)\). Since one lasso contributes only \(O(N^{2/3})\) pieces, a construction with superlinearly many pieces must use unboundedly many lassos. The second concerns a monitor or controller that observes only the received outputs instead of the graph vertex; this leads to games with imperfect information, whose burst boundaries we have not characterized.

\bibliographystyle{alphaurl}
\bibliography{rate_burst}

\end{document}